\documentclass[12pt]{article}
\pdfoutput=1

\usepackage{amsfonts}
\usepackage{amssymb}
 
\usepackage{hyperref}
\newlength{\abstractwidth}
\usepackage{color}
\usepackage{graphicx}

\renewcommand{\thefootnote}{\fnsymbol{footnote}}
\renewcommand{\thanks}[1]{\footnote{#1}}
\newcommand{\starttext}{
\setcounter{footnote}{0}
\renewcommand{\thefootnote}{\arabic{footnote}}}

\newcommand{\bea}{\begin{eqnarray}}
\newcommand{\eea}{\end{eqnarray}}
\newcommand{\ee}{\end{equation}}
\newcommand{\be}{\begin{equation}}

\newcommand{\no}{\nonumber}

\def\cA{{\cal A}}

\def\cD{{\cal D}}
\def\cE{{\cal E}}

\def\cK{{\cal K}}

\def\cM{{\cal M}}
\def\cN{{\cal N}}
\def\cO{{\cal O}}

\def\mN{\mathfrak{N}}

\def\mg{\mathfrak{g}}

\def\bs{{\bf s}}

\def\det{{\rm det}}

\def\half{ {1\over 2}}
\def\p{\partial}

\def\b{\beta}

\def\ep{\varepsilon}

\def\g{\gamma}

\def\om{\omega}

\def\ZZ{{\mathbb Z}}
\def\RR{{\mathbb R}}

\def\CC{{\mathbb C}}

\def\no{\nonumber}
\def\sm{\smallskip}

\long\def\symbolfootnote[#1]#2{\begingroup%
\def\thefootnote{\fnsymbol{footnote}}\footnote[#1]{#2}\endgroup}

\begin{document}
\starttext
\setcounter{footnote}{0}

\begin{flushright}
19 June 2014
\end{flushright}

\bigskip

\begin{center}

{\Large \bf  Holographic Entropy and Calabi's Diastasis}

\vskip 0.4in

{\large  Eric D'Hoker and Michael Gutperle}

\vskip .2in

{ \sl Department of Physics and Astronomy }\\
{\sl University of California, Los Angeles, CA 90095, USA}\\
{\tt  dhoker@physics.ucla.edu;    gutperle@physics.ucla.edu }

\end{center}

\begin{abstract}

\vskip 0.1in

The entanglement entropy for interfaces and junctions of two-dimensional CFTs is evaluated 
on holographically dual half-BPS solutions to six-dimensional  Type 4b supergravity with $m$
anti-symmetric tensor supermultiplets. It is shown that the moduli space for an $N$-junction 
solution projects to $N$ points in the K\"ahler manifold $SO(2,m)/\left ( SO(2) \times SO(m) \right )$.
For $N=2$ the interface entropy is expressed in terms of the central charge and Calabi's diastasis 
function on $SO(2,m)/\left ( SO(2) \times SO(m) \right )$,
thereby lending support from holography to a proposal of Bachas, Brunner, Douglas, and Rastelli.
For $N=3$, the entanglement entropy for a 3-junction decomposes into a sum of diastasis functions 
between pairs, weighed by combinations of the three central charges,  provided the flux charges
are all parallel to one another or, more generally, provided the space of flux charges is 
orthogonal to the space of unattracted scalars. 
Under similar assumptions for $N \geq 4$, the entanglement entropy for the $N$-junction  
solves a variational problem whose data consist of the $N$ central charges,
and the diastasis function evaluated between pairs of $N$ asymptotic $AdS_3 \times S^3$ regions.

\end{abstract}

\baselineskip=12pt
\setcounter{equation}{0}
\setcounter{footnote}{0}

%
%
%
%
%
\newpage

\baselineskip 16pt 

\section{Introduction}
\setcounter{equation}{0}
\label{sec1}

An increasingly compelling connection has been emerging between entropy and geometry
ever since Bekenstein and Hawking assigned an entropy to a quantum black hole.
Gauge/gravity duality  \cite{Maldacena:1997re,Gubser:1998bc,Witten:1998qj} relates 
the entropy of a thermal state in quantum field theory
to the entropy of a black holes in the gravity dual. The Ryu-Takayanagi 
proposal \cite{Ryu:2006bv,Ryu:2006ef}  gives a holographic formula for the entanglement 
entropy in conformal field theory 
associated with a spatial region $\cA$ in terms of the area of a minimal surface in the bulk 
gravity theory subtended by the region $\cA$. For conformal field theories with interfaces
or boundaries, the entanglement entropy provides information on the degeneracy of the 
ground state $\mg$-function, also referred to as the interface or boundary entropy \cite{Affleck:1991tk}. 

\sm

Superconformal gauge theories in which an interface, a defect,  or a boundary is preserved 
by part of the superconformal symmetry have been the subject of intense study, in large
part because these theories often provide solvable yet non-trivial deformations of the 
original theory. Such studies include probe brane constructions \cite{Karch:2000gx}; the construction of two-dimensional 
conformal interfaces by the folding trick \cite{Bachas:2001vj}; the discovery of topological 
defects  and their algebra \cite{Bachas:2007td,Fuchs:2007tx}; the analysis of 
supersymmetry preserving interfaces in four-dimensional $\cN=4$ super-Yang-Mills \cite{Clark:2004sb,D'Hoker:2006uv,Gaiotto:2008sa,Gaiotto:2008sd}; 
and the interplay between defect and domain wall operators \cite{Gaiotto:2009fs,Drukker:2010jp}. 
Rich families of supersymmetric fully back-reacted solutions have been constructed in Type IIB 
supergravity for supersymmetric interfaces in \cite{Clark:2005te,D'Hoker:2007xy} and 
Wilson lines in \cite{Lunin:2006xr,D'Hoker:2007fq}; in M-theory for defects   in 
\cite{D'Hoker:2008wc,D'Hoker:2007xz,Lunin:2007ab}; and in various supergravities
for junctions of CFTs in two dimensions in 
\cite{Chiodaroli:2009xh,Chiodaroli:2010mv,Chiodaroli:2011nr,Chiodaroli:2011fn}.

\sm

A intriguing novel connection was proposed in \cite{Bachas:2013nxa} between the interface 
entropy in certain two-dimensional CFTs and Calabi's diastasis function of K\"ahler geometry. 
In string theory, K\"ahler geometry governs compactifications which preserve various degrees
of space-time supersymmetry.  The moduli spaces of the corresponding $(2,2)$ supersymmetric 
sigma models generically have K\"ahler moduli and complex structure moduli components. 

\sm

The Calabi diastasis function \cite{calabi} may be defined for any K\"ahler manifold $\cK$ with 
K\"ahler form $\om= i \p \bar \p K$ and associated K\"ahler potential $K$. The K\"ahler form $\om$
is invariant under K\"ahler gauge transformations, which may be expressed in local complex 
coordinates $(t,\bar t)$ by $K(t,\bar t) \to K(t,\bar t) + \Lambda (t) + \bar \Lambda (\bar t)$,
where $\Lambda (t)$ is holomorphic. Calabi showed \cite{calabi} that the real-valued K\"ahler 
potential $K(t,\bar t)$ may be continued to a complex-valued potential $K(t_1, \bar t_2)$ for 
independent points $t_1$ and $t_2$. The diastasis function, 
\bea
\label{diadef}
\cD (1,2)= K(t_1,\bar t_1)+ K(t_2,\bar t_2)- K(t_1,\bar t_2) - K(t_2,\bar t_1)
\eea
is then well-defined, invariant under K\"ahler gauge transformations, and preserved upon
restriction to a complex analytic submanifold of $\cK$. In the limit where the points 
$t_1,t_2$ are infinitesimally near one another, $\cD(1,2)$ reduces to the K\"ahler metric on $\cK$.

\sm

Specifically, a formula was proposed in \cite{Bachas:2013nxa} for the $\mg$-function of an interface 
separating $(2,2)$ supersymmetric CFTs with K\"ahler moduli  $t_1$ and $t_2$ 
in terms of the diastasis function, 
\bea
\label{diarela}
2 \ln \mg = \cD (1,2)
\eea
In turn, the $\mg$-function is related to the entanglement entropy of a spatial  region of length $L$ 
which encloses the interface symmetrically, by the following relation \cite{Azeyanagi:2007qj}, 
\bea
S_\ep= {c\over 3} \ln {L\over \ep}+  \ln \mg
\eea
The examples given in \cite{Bachas:2013nxa} to illustrate the relation (\ref{diarela}) include  
sigma models with $(2,2)$ supersymmetry, for target space $T^2$ as well as Calabi-Yau 
manifolds in the large volume limit. 
A common feature of these examples is the fact that (\ref{diarela}) holds only for a special subclass of interfaces which preserve some supersymmetry, and for interfaces where the moduli of either the 
complex structure or the K\"ahler  structure are held fixed across the interface. 

\subsection{Summary of results}

In the  present paper we shall produce evidence supporting the relation (\ref{diarela})
between the interface entropy and Calabi's diastasis function using the holographically dual
half-BPS interface solutions \cite{Chiodaroli:2011nr} to Type 4b supergravity \cite{Romans:1986er}.
The fundamental property of these families of  solutions that makes this correspondence possible
is the existence of a smooth projection from their $(3m+2)$-dimensional moduli space 
to a pair of points in the K\"ahler manifold $SO(2,m)/\left ( SO(2)\times SO(m) \right )$.
The points correspond to the two asymptotic $AdS_3 \times S^3$ regions of the interface,
and are subject to the overall conservation of anti-symmetric tensor field flux charge. The interface entropy 
is then determined by  the Calabi diastasis function evaluated at this pair of points, along with 
the common central charge of these regions.  Essential in making this connection is the
fact that the holographic interface solution preserves some supersymmetry.  
Note that supersymmetry was also a crucial ingredient on the CFT side, as discussed in 
\cite{Bachas:2013nxa}. There is was shown that the relation of $\mg$-function and diastasis function 
does not hold  for  a non-supersymmetric interface where both the  K\"ahler  and complex 
moduli jump. A holographic example of this failure is given in section \ref{sec8}, where it is 
shown that for a nonsupersymmetric Janus interface the $\mg$-function is related to the geodesic distance between points in the moduli space rather than to the diastasis function.

\sm

Next, we shall define and evaluate the entanglement entropy of the half-BPS 
solutions to Type 4b supergravity which are dual to $N$-junctions.  
The corresponding $N$-junctions solutions 
were obtained explicitly in \cite{Chiodaroli:2011nr}. Their space-time manifold is of 
the form $AdS_2 \times S^2$ warped over a Riemann surface $\Sigma$ with boundary $\p \Sigma$. 
The solutions have $N$ asymptotic $AdS_3 \times S^3$ regions labelled by $i=1,\cdots, N$, 
each of which is characterized by a unit vector $\hat \kappa_i \in \RR^{m+2}$ of vacuum 
expectation values of the un-attracted scalars, as well as a charge vector $\mu_i \in \RR^{m+2}$ 
which obeys\footnote{The dot product stands for the $SO(2,m)$-invariant inner product 
with signature $(++ - \cdots -)$.} $\mu_i \cdot \mu_i >0$, overall charge conservation 
$\sum _{i=1}^N \mu_i=0$, as well as $\mu_i \cdot \hat \kappa _i =0$.  The data 
$\hat \kappa _i, \mu_i$, subject to the above relations, account for the 
$2(m+1)N-m-2$ moduli of these families of solutions, including for the central
charge $c_i \sim \mu_i \cdot \mu_i$ of each asymptotic  region.

\sm

The supergravity fields of the general half-BPS $N$-junction solutions are completely determined 
in terms of  $\hat \kappa _i$ and $\mu_i$ by the BPS equations and Bianchi identities \cite{Chiodaroli:2011nr}.
We shall prove a key result that all data are equivalently and uniquely determined by 
extremizing the holographic entanglement entropy for given $\hat \kappa _i$ and $\mu_i$. 
This result may be interpreted as a realization (albeit in a ``mini-superspace" sense) of the idea 
that gravitational equations of motion follow from 
entanglement entropy (see e.g. \cite{Lashkari:2013koa,Faulkner:2013ica})

\sm

Finally, we shall derive generalizations applicable to the entanglement entropy of the 
junctions of $N \geq 3$ CFTs, each of which lives on a spatial half-line, and which
are joined at a single spatial point. We shall often refer to the entanglement entropy in 
this case as {\sl junction entropy}, and derive a general formula for the junction entropy 
of all such solutions in terms of the data $\hat \kappa _i$ and $\mu_i$ for $i=1,\cdots, N$. 
For special arrangements of the charges, such as when all charge
vectors being parallel to one another,  we shall express the junction entropy as a
sum of terms each of which is governed by the diastasis function for a pair
of $AdS_3 \times S^3$ regions. We end by speculating on the significance of the 
junction entropy as an $N$-point generalization of the diastasis function with $N \geq 3$.
We shall also briefly discuss the possible significance of the special
arrangements of charges upon which the junction entropy reduces to a dependence on
Calabi's diastasis function only.

\subsection{Organization}

The remainder of this paper is organized as follows. In section \ref{sec2} we review  
the six-dimensional Type 4b supergravity solutions which are half-BPS and describe 
holographic interfaces and junctions.  In section \ref{sec3} we calculate the entanglement 
entropy for the general $N$-junction solution.  In section \ref{sec4} we analyze the 
K\"ahler structure of the moduli space of half-BPS solutions and express the diastasis 
function in terms of the supergravity fields.  In section \ref{sec5}  we give a holographic
proof of the relation (\ref{diarela}) between the  interface entropy and the diastasis function  
for $N=2$.
In section \ref{sec6} we calculate the entanglement entropy of the holographic solutions which 
are dual to  junctions of three CFTs. Specializing to the case of parallel charges $\mu _i$,
or when $\hat \kappa _i \cdot \mu_j=0$ for all $i,j$, we express the junction 
entropy as a sum of  diastasis functions of pairs of asymptotic data. 
In section \ref{sec7} we present an analogous treatment for the case of  $N$-junctions. 
We close in section \ref{sec8}  with a calculation of the entropy for a non-supersymmetric 
interface, and in section \ref{sec9} with a discussion of our results and future directions. 
Some review material and technical details of the UV regularization of the holographic  
entropy are relegated to Appendix \ref{appb}.


\section{Holographic Interfaces and Junctions}
\setcounter{equation}{0}
\label{sec2}

The holographic dual to two-dimensional CFTs with an interface or a junction 
will be formulated in terms of six-dimensional Type 4b supergravity \cite{Romans:1986er},
a family of theories which contain a supergravity supermultiplet and $m$ anti-symmetric tensor supermultiplets.  
The bosonic fields consist of the metric, two-form fields of which 5 have self-dual and $m$ have 
anti-self dual field strength,  and $5m$ scalars  in the $SO(5,m)/\left ( SO(5)\times SO(m) \right )$
coset. The fermionic fields consist of four negative chirality gravitinos and $4m$ positive chirality 
spinors. Classically, the number $m$ is arbitrary and the supergravity Bianchi identities
and field equations are invariant under $SO(5,m,\RR)$. At the quantum level, however, the absence 
of anomalies requires  $m=5$ or $m=21$, and restricts invariance to the U-duality group $SO(5,m,\ZZ)$.
The theory corresponds to the low energy limit of Type IIB string theory compactified 
respectively on the spaces $T^4$ or $K3$.

\sm

The vacuum solution has space-time $AdS_3 \times S^3$ and is invariant under the 
isometry  $PSU(1,1|2)\times PSU(1,1|2)$ Lie superalgebra. The dual CFT has a central
charge related to the radius of $AdS_3$ by the Brown-Henneaux formula \cite{Brown:1986nw}.

\sm

A half-BPS solution which is holographically dual to the interface of two CFTs interpolates between two 
asymptotic $AdS_3 \times S^3$ regions with the same central charge.  A half-BPS solution dual  to the junction of $N$ different CFTs is characterized by a space-time with 
$N$ asymptotic $AdS_3 \times S^3$ regions, in which the radii of the asymptotic $AdS_3$
are subject to certain mild inequalities. Regular solutions to Type 4b 
supergravity with these properties exist for arbitrary $m$ 
and have been constructed explicitly in \cite{Chiodaroli:2011nr,Chiodaroli:2010ur}.

\subsection{Half-BPS supergravity solutions in Type 4b}

In this section we shall briefly review the salient features of the half-BPS solutions to Type 4b 
supergravity of \cite{Chiodaroli:2011nr,Chiodaroli:2010ur} which are dual to 
interface and junction CFTs. The structure of their space-time manifold is dictated by 
supersymmetry. It takes the form of $AdS_2 \times S^2$ warped over a Riemann
surface $\Sigma$ with boundary and enjoys  a $SO(2,1)\times SO(3)$ isometry.  

\sm

The space-time metric $ds^2$ of the solutions and its closed 3-form field strengths
$G^A$ with $A=1,\cdots, m+5$,  are given as follows, 
\bea
\label{2a1}
ds^2 & = & H{F_+ \over F_-}  \, ds_{AdS_2}^2 
+ H { F_-\over F_+}  \, ds^2 _{S^2} + {F_+ F_- \over H}  |dw|^2
\no \\
G^A & = & d\Psi ^A \wedge \om _{AdS_2} + d\Phi ^A \wedge \om_{S^2}
\eea
Here, $w,\bar w$ are local complex coordinates on $\Sigma$,
while $ds^2_{AdS_2}$ and $ds^2 _{S^2}$ are the metrics respectively of the manifolds 
$AdS_2$ and $S^2$ with unit radius, and $\om _{AdS_2}$ and $\om _{S^2}$ are their 
respective volume forms. The remaining data, namely $H$, $F_\pm$, $\Phi^A$ and $\Psi ^A$
are all real-valued functions on $\Sigma$, which we shall now specify.\footnote{The data used
in the notations of \cite{Chiodaroli:2011nr} are related to the data used here by 
 $f_1^2=HF_+/F_-$, $f_2^2 = H F_-/F_+$, and $4\rho^2= F_+ F_-/H$.}

\sm

The BPS equations and regularity conditions require $H$ to be a positive 
harmonic  function in the interior of $\Sigma$ which vanishes on the boundary $\p \Sigma$ of $\Sigma$.
They also require $F_\pm^2$ to be positive in the interior of $\Sigma$ and $F_-$
to vanish on $\p \Sigma$. The BPS equations, along with the Bianchi identities,
then determine the remaining data in terms of an $SO(5,m)$-vector $\Lambda $ of  
meromorphic functions on $\Sigma$ satisfying,
\bea
\label{2a2}
0 & = & \Lambda \cdot  \Lambda  - 2 (\p_w H)^2 
\no \\
F_\pm^2 & = & \bar \Lambda \cdot  \Lambda  \pm 2 |\p_w H|^2
\eea 
The dot product is taken with respect to the $SO(5,m)$-invariant metric 
$\eta = \hbox{diag} (I_5, - I_m)$. The real-valued flux potentials $\Phi^A$ and $\Psi ^A$ 
are  given in terms of the complex combination, 
\bea
\label{2a3}
{1 \over \sqrt{2}} \left ( \Phi ^A - i \Psi ^A \right ) =
{ H \p_w H \over F_+^2 F_-^2} 
\Big ( (\bar \Lambda \cdot \bar \Lambda) \, \Lambda ^A 
- (\bar \Lambda \cdot \Lambda )\, \bar \Lambda ^A \Big )
- \int dw \, \Lambda ^A 
\eea
It is a fundamental result, obtained in \cite{Chiodaroli:2011nr}, that the BPS equations  
require invariance of half-BPS solutions 
under an $SO(3)$ subgroup of the $SO(5)$ factor of $SO(5)\times SO(m)$.
The $SO(3)$ is minimal in $SO(5)$ so that the vector of $SO(5)$ 
decomposes under $SO(3)$ as follows ${\bf 5} = {\bf 3} \oplus {\bf 1} \oplus {\bf 1}$. 
As a result, the invariance of any $SO(5,m)$ vector under this 
$SO(3)$ requires the vanishing of the corresponding components of the vector. We shall choose 
a gauge in which $G^A=\Phi^A=\Psi ^A=\Lambda ^A=0$ for $A=3,4,5$. 
 
 \sm
 
Finally, the matrix of scalar fields $V$ takes values in $SO(5,m)/ \left ( SO(5)\times SO(m) \right )$, 
so that we have $V^t \eta V=\eta$, where $\eta = \hbox{diag} (I_5, -I_m)$. We shall
denote its components by $V= V^{(\rho ,r)} {}_A$ with $\rho=1,\cdots, 5$ and $r=6, \cdots, m+5$.
The $SO(3)$ invariance of the solutions implies that we should set 
$V^{(\rho ,r)}{}_A=\delta ^\rho {}_A$ for $A=3,4,5$ as well as for $\rho =3,4,5$.
The remaining $V$ effectively takes values in the reduced space 
$SO(2,m)/\left ( SO(2)\times SO(m) \right )$,
and may be parametrized, uniquely up to rotations in $SO(2) \times SO(m)$, by the entries,
\bea
\label{2a4}
V^\pm {}_A = { 1 \over \sqrt{2} } \left ( V^1 {}_A \pm i V^2 {}_A \right )
\hskip 1in 
V^+ {}_A =  \left ( \bar \lambda _A - |X|^2 \lambda _A \right )\, X \, (1-|X|^4)^{-1}
\eea
with $|X|^2 + |X|^{-2} = \bar \Lambda \cdot \Lambda / |\p_w H|^2$. In particular, 
the phase of the function $X$ remains undetermined, as it transforms non-trivially
under $SO(2)$ rotations in $SO(2) \times SO(m)$.

\sm

Note that the effective target space $SO(2,m)/\left ( SO(2)\times SO(m) \right )$ of the scalar fields 
is the K\"ahler manifold which will govern the Calabi diastasis structure to be established below.

\subsection{Parametrization of the supergravity solutions}

We limit attention here to the case where the Riemann surface $\Sigma$ has only a single 
connected boundary component and no handles, so that it may be modeled by the upper half plane.\footnote{Generalizations to Riemann surfaces with multiple boundaries and handles 
were discussed in \cite{Chiodaroli:2012vc}.} Positivity of $H$ and $F_\pm^2$  in the 
interior of $\Sigma$ forces all poles $x_i$ of the harmonic function $H$ to lie 
on the real line, and $\Lambda$ to have single and double poles at $x_i$. Regularity of the  solution 
precludes $\Lambda$ from having singularities away from the points $x_i$.

\sm

Near each pole  $x_i$ the metric becomes locally asymptotic to $AdS_3\times S^3$
and corresponds to a half-line CFT holographic dual. Thus a supergravity solution with 
$N$ poles $x_i$ for $i=1,\cdots, N$ will produce a holographic dual consisting of a junction 
of $N$ half-line CFTs.
The basic functions of these solution take the following form, 
\bea
\label{2b1}
H&=& i \sum_{i=1}^N \left ( {\gamma_i\over w-x_i} - { \gamma_i\over \bar w-x_i} \right )
\no \\
\Lambda^A &=&  -i  \sum_{i=1}^N \left ( {  \kappa_i^A \over( w-x_i)^2 }  
+ { \mu_i^A\over w-x_i} \right ) 
\eea
The residues $\gamma _i , \kappa _i ^A$, and $\mu_i^A$ are real,
with $\gamma _i >0$. The  index $A$ ranges over $A=1,\cdots, m+5$
with the understanding that $SO(3)$ invariance sets 
$\Lambda ^A=\kappa _i ^A=\mu_i^A=0$ for $A=3,4,5$.

\sm
 
The residue $\mu_i^A$ gives the charge (or flux) of the 3-form field strength $G^A$
across a three-sphere $S_i^3$ in the asymptotic $AdS_3 \times S^3$ region 
at the pole $x_i$. Using (\ref{2a3}) we find,
\bea
\label{2b2}
2 \sqrt{2} \pi ^2 \mu _i ^A=  \int _{S^3 _i} G^A
\hskip 1in 
\sum_{i=1}^N\mu_i^A=0
\eea
The second equation above expresses  overall charge  conservation. 
The first equation of (\ref{2a2}) is equivalent to the following constraints for each $i=1,2,\cdots,N$,
 \bea
\label{2b3}
\kappa _i \cdot \kappa _i = 2 \gamma_i^2  & \hskip 1in & \cE _i ^{(1)} =0
\no \\
\kappa _i \cdot \mu_i=0 ~~~ &&   \cE_i ^{(2)}=0
\eea
where $\cE_i ^{(1)}$ and $\cE_i^{(2)}$ are given as follows,
 \bea
\label{2b4}
\cE_i ^{(2)} 
& = & 
\mu _i \cdot \mu _i  +
2 \sum _{j \not= i} {\kappa _i \cdot \mu  _j \over x_i - x_j} 
- 2 \sum _{j \not=i} { 2\gamma_i \gamma_j - \kappa  _i \cdot \kappa  _j \over (x_i-x_j)^2} 
\no \\
\cE_i ^{(1)} 
& = &  
 \sum _{j \not= i} { \mu _i \cdot   \mu _j \over x_i - x_j}
- \sum _{j \not= i} { \kappa _i \cdot \mu  _j - \mu _i \cdot \kappa _j  \over (x_i - x_j)^2} 
+ 2 \sum _{j \not=i} { 2\gamma_i \gamma_j -  \kappa  _i \cdot \kappa _j \over (x_i -x_j)^3} 
\eea
For $N\geq 3$, there exist three relations between $\cE_i^{(1)}$ and $\cE^{(2)}_i$, namely
for $n=0,1,2$, we have,
\bea
\sum _{i=1}^N \left (  x_i^n \, \cE_i ^{(1)} - n  x_i^{n-1} \cE^{(2)}_i \right )=0
\eea
Thus, the equations $\cE^{(1)}_i=\cE_i^{(2)}=0$ constitute $2N-3$ independent constraints.
It is straightforward to verify the $SL(2,\RR)$-covariance of equations (\ref{2b3})
under which $\mu_i$ is invariant while the other data transform as follows,
\bea
x_i ' = { a x_i + b \over c x_i +d} 
\hskip 1in 
\gamma_i ' = { \gamma _i \over ( cx_i +d)^2}
\hskip 1in 
\kappa_i ' = { \kappa _i \over ( cx_i +d)^2}
\eea
with $a,b,c,d \in \RR$ and $ad-bc=1$. The data $x_i, \gamma _i, \kappa _i, \mu_i$ in the 
functions $H$ and $\Lambda^A$ for the $N$-junction solution  contain $2N(m+3)-3$ 
real parameters, taking into account that we set $\kappa _i^A=\mu_i^A=0$ for $A=3,4,5$,
as well as the covariance under $SL(2,\RR)$.  The number of charge conservation relations in (\ref{2b2}) 
is $m+2$, while the number of independent constraints in (\ref{2b3}) is $4N-3$, leaving 
$2N(m+1)-(m+2)$ independent moduli.

\subsection{Asymptotic $AdS_3\times S^3$ regions}

To analyze the asymptotic behavior of the metric of the solutions, given in (\ref{2a1}), 
we begin by parametrizing the $AdS_2$ factor in terms of the unit radius Poincar\'e patch metric,
\be
\label{ads2met}
ds_{AdS_2}^2= {dz^2-dt^2\over z^2}
\ee
where $t \in \RR$ denotes time, and $z \in \RR^+$.
Near the poles of the harmonic function $H$ the metric becomes locally asymptotic to $AdS_3\times S^3$. 
The asymptotic behavior can be exhibited   by defining  $w=x_i + e^{-x+i \theta}$ 
and expanding the metric functions in the limit $x \to + \infty $,
\bea
\label{asymmet}
ds^2 =  \sqrt{2 \mu_i \cdot \mu_i} 
\left ( { dx^2} + {8 \gamma_i^2 \over \mu _i \cdot \mu_i } \, e^{2x} \, ds^2 _{AdS_2}
+ d \theta ^2 + \sin^2 \theta \, ds^2 _{S^2} \right )+ \cO (e^{-2x})
\eea
Since the $AdS_2$ factor in (\ref{asymmet}) is conformal to the half line $\times$ time,  
the conformal boundary of the metric contains $N$ half-spaces, parameterized by $t,z$ 
which are glued together at the boundary of $AdS_2$ located at $z=0$. 
Hence the holographic interpretation of the solution is that of an $N$-junction where $N$ 
different CFTs, each of which is defined on $\RR^+$, are glued together at a one-dimensional 
junction. It follows from (\ref{asymmet}) that the radius $R_i$ of the $i$-th asymptotic $AdS_3$  
region, and hence the central charge $c_i$ of the dual CFT, are given by,
\bea
\label{2c3}
R_i^4 = 2 \mu_i \cdot \mu_i \hskip 1in c_i = { 6 \pi^2 \mu_i \cdot \mu_i \over  G_N}
\eea
where $G_N$ is the six dimensional  Newton's constant. The scalar fields of $V$ lie in a
the K\"ahler coset space $SO(2,m)/\left ( SO(2)\times SO(m) \right )$. 
In the $i$-th  asymptotic region the scalars have the following limiting behavior, 
\be
\label{asymscal}
V^\pm {}_A (x_i) 
=  e^{i\theta_i} \, 
\left ( {\kappa_{iA} \over 2 \gamma_i} \pm i {\mu_{iA} \over \sqrt{2 \mu_i \cdot \mu_i} }\right )
\ee
where $\theta_i$ is the phase of the field $X$ of (\ref{2a4}) at the pole $x_i$.  
Note that the second term in (\ref{asymscal}) is completely determined by  the charges 
$\mu_i^A$ in the $i$-th asymptotic   $AdS_3$ region. Consequently, these scalars are 
subject to an attractor mechanism. By contrast, the first term in (\ref{asymscal}) 
is not fixed by the charges and the corresponding scalars are un-attracted.

\subsection{Supergravity solutions dual to interfaces and junctions}

There is no regular solution with $N=1$, although relaxing the regularity conditions allows 
for (singular) holographic duals of boundary CFT with only one asymptotic AdS 
region \cite{Chiodaroli:2011fn}.

\sm

Firstly, we will consider $N=2$ regular solutions with two asymptotic 
$AdS_3 \times S^3$ regions. They are holographically dual to a half-BPS interface CFT.  
Charge conservation requires the CFTs on both sides have the same central charge, 
but the values of  un-attracted scalars may jump across the interface. The $N=2$ solution 
is therefore a realization of a Janus configuration \cite{Bak:2003jk} in six dimensional supergravity.

\sm

Secondly, we consider $N=3$ regular solutions with three asymptotic $AdS_3 \times S^3$ regions.
They are holographically dual to a junction of three CFTs. In this case charge conservation 
allows the three CFTs which meet at the junction to have different  central charges,
and correspond to a decoupling limit of different self-dual strings in six dimensions.

\sm

Thirdly, we consider  $N$-junction regular solutions with $N \geq 4$ asymptotic $AdS_3 \times S^3$ regions,
which are holographically dual to $N$ different  CFTs  meeting at one point. Solving the constraints 
of (\ref{2b3}) and (\ref{2b4}) is now considerably more involved than for 3-junctions and interfaces,
and no closed-form analytical solution is known at this time.

\section{Entanglement Entropy}
\setcounter{equation}{0}
\label{sec3}

In this section we shall calculate the entanglement entropy for the half-BPS interface and junction 
solutions and extract the boundary entropy (or g-function) from the results. In particular we shall 
discuss the required careful regularization of the integrals involved and the cutoff dependence of the result. 
The  connection to the diastasis function for $N=2$, $N=3$, and $N>3$ will be made in 
sections \ref{sec5}-\ref{sec7}.

\begin{figure}[hb]
\centering
\begin{tabular}{c}
\includegraphics[width=75mm]{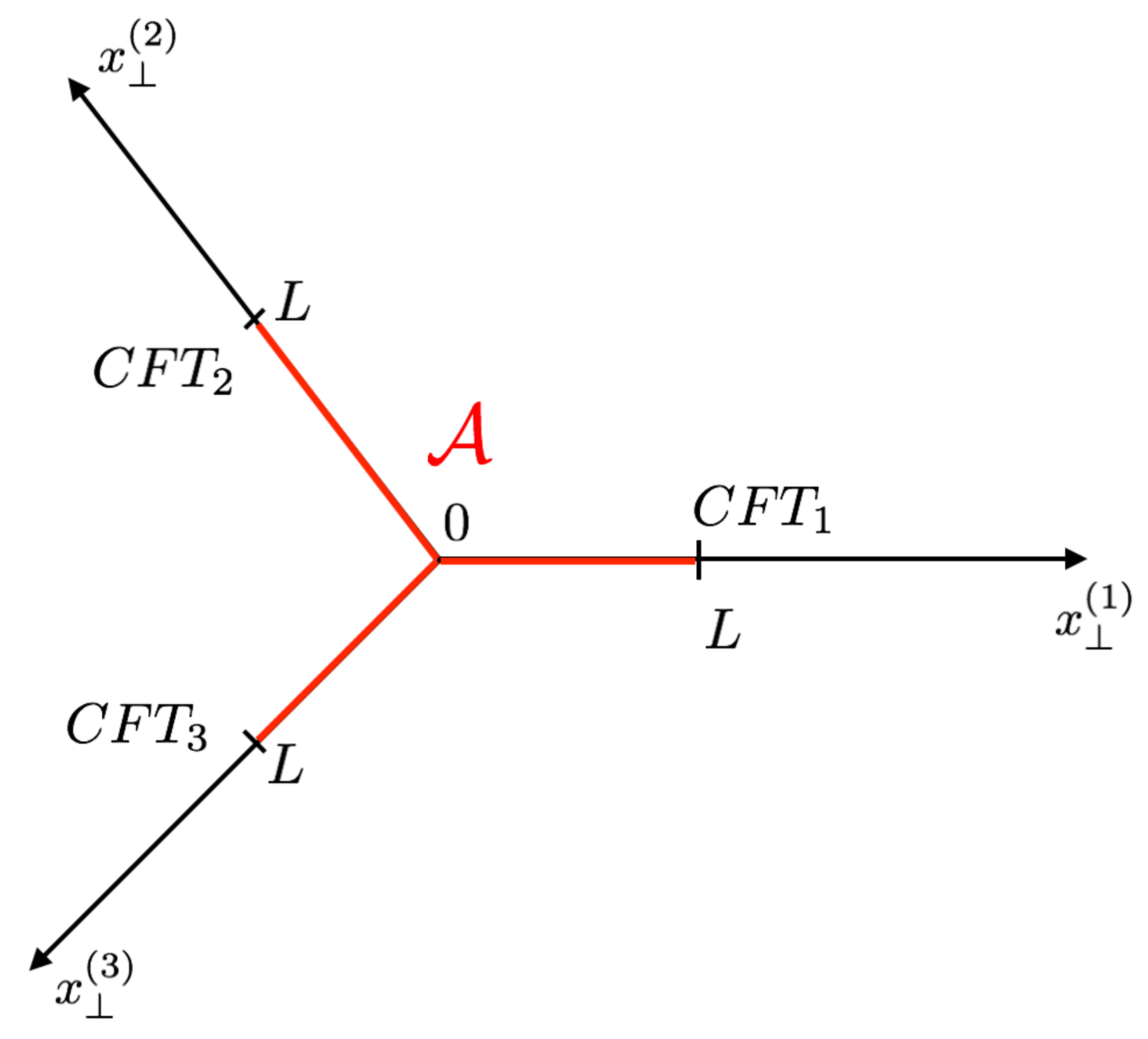}
\end{tabular}
\caption{ Junction of three CFTs and star shaped entangling surface $\cA$. }
\label{fig1a}
\end{figure}

We choose the entangling region $\cA$  to enclose the interface symmetrically. 
For the $N$-junctions, we choose a symmetric star-shaped region  (see figure \ref{fig1a} 
for an example of a $N=3$ junction) which extends the same distance in all 
half-spaces.   The Ryu-Takayanagi prescription \cite{Ryu:2006bv} states that the 
entanglement entropy is given by the area of a minimal surface in the bulk which encloses  the 
boundary $\partial \cA$ of the region $\cA$ when it reaches the  asymptotic AdS  boundary. 
This prescription works straightforwardly for three dimensional spacetimes which asymptote to 
$AdS_3$. For the BPS junctions we have to generalize the prescription since the solution is a fibration of $AdS_2\times S^2$ over the upper half plane $\Sigma$.  The minimal area surface 
for the holographic entanglement entropy is given by holding the $AdS_2$ time  $t$ 
constant, setting $z=L$, and integrating over the two sphere $S^2$ and  the Riemann surface $\Sigma$
(see figure 2).
The entanglement entropy is then given by the area of this surface, and its expression may
be read off using the metric of (\ref{2a1}),
\bea
\label{3a1}
S_e = {1\over {4G_N} }\int_{\Sigma }|dw|^2 F_-^2 \int_{S^2}  \om_{S^2} 
=  { \pi \over  G_N} \int _{\Sigma } |dw|^2 \left ( \bar \Lambda \cdot \Lambda  
- 2 |\p_w H|^2 \right )
\eea
Here, $G_N$ is Newton's constant.
The second formula is obtained by integrating over $S^2$ in the first formula,  
and using the expressions for $F_-^2$  given in (\ref{2a2}). 

\sm

The above formula for the entanglement entropy $S_e$ is formal, as the integration over $\Sigma$ 
diverges due to the presence of the poles of $H$ and $\Lambda ^A$ at the boundary of $\Sigma$. 
To regularize these divergences, we introduce a cutoff by removing a (half-) disk of coordinate radius 
$\ep_i >0$ around the pole $x_i$ for all $i=1,\cdots, N$ (see figure 2). 

\begin{figure}[hb] 
\centering
\begin{tabular}{c}
\includegraphics[scale=0.9]{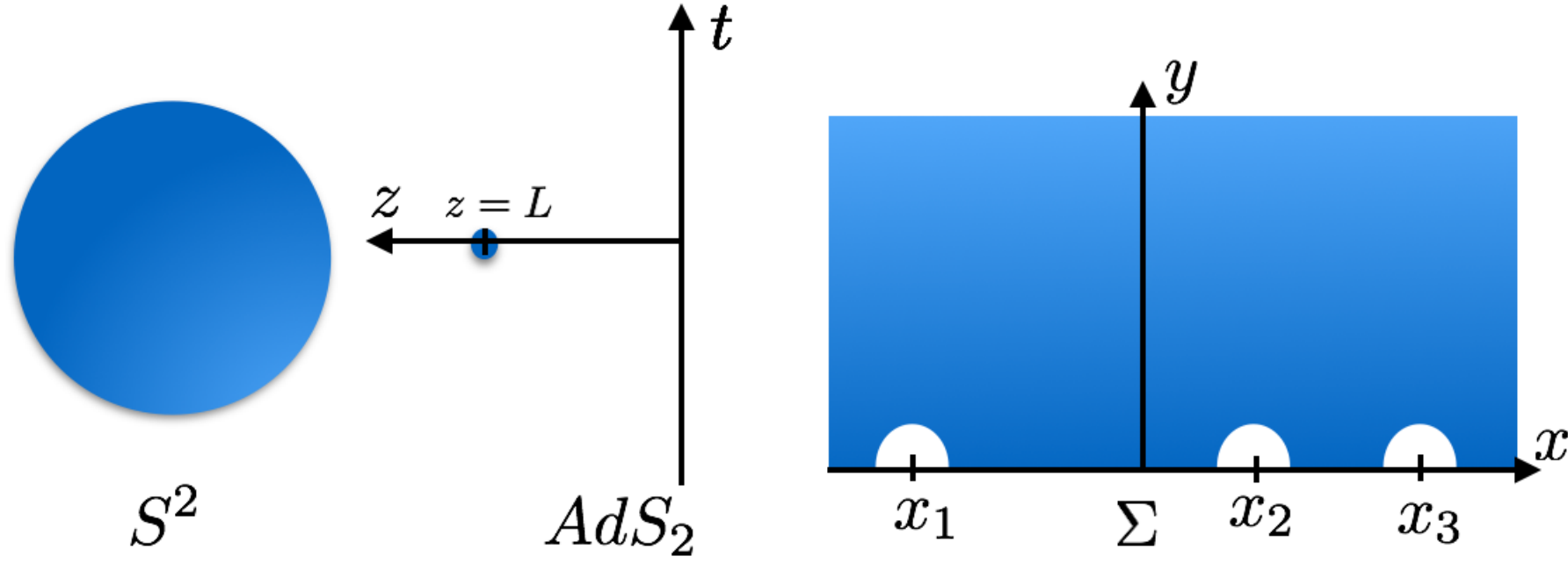}
\end{tabular}
\caption{ The $AdS_2\times S^2 \times \Sigma$ space-time for $N=3$ junctions. 
The surface of minimal area is localized in $AdS_2$  at $z=L$, and encompasses 
all of $S^2 \times \Sigma$.}
\label{fig2}
\end{figure}

Within the context of AdS/CFT, the 
cutoffs $\ep_i$ at different poles must be related to the common UV-cutoff $\ep$ of the dual 
CFT. In Appendix \ref{appb}, we shall provide a careful derivation
of the corresponding relation, 
\bea
\label{3a2}
\ep_i ^2 =  { 4 \kappa_i \cdot \kappa_i  \over  \mu_i  \cdot \mu_i} \,  {\ep^2\over L^2} 
\eea
using the Fefferman-Graham expansion.

\sm

Convergence of the integral in (\ref{3a1}) for large $|w|$ is guaranteed by the flux conservation 
formula of (\ref{2b2}). Still, to evaluate the integrals of individual terms in (\ref{3a1}) arising from the substitution of (\ref{2b1}), it is convenient to also introduce a large $|w|$-cutoff $W$ so that $0<|w|<W$. 
The key integral needed to evaluate (\ref{3a1}) is then given by, 
\bea
\label{3a3}
\int _{\Sigma } { |dw|^2 \over (w-x)(\bar w -y)} = 
{ \pi \over 2} \ln \left ( { W^2 \over (x-y)^2 + \ep^2} \right )
\eea
along with its derivatives in $x$ and/or $y$.
The result for the entanglement entropy becomes,
\bea
\label{ent1}
S_e & = & 
 { \pi^2 \over  G_N} \, S_R 
+{ \pi ^2 \over 2G_N} \sum _{i=1} ^N \mu _i \cdot \mu_i  
\ln \left ( { \mu_i \cdot \mu_i  L^2\over 8  \ep ^2} \right )
\\
S_R & = & \sum _{i=1}^N \sum _{j\not= i} \left ( 
{ 2 \gamma_i \gamma_j - \kappa _i \cdot \kappa _j \over (x_i -x_j)^2} 
+ { \kappa _j \cdot \mu_i - \kappa _i \cdot \mu _j \over x_i - x_j}
- \half \mu_i \cdot \mu_j \ln { (x_i -x_j)^2 \over \gamma _i \gamma _j} \right )
\no
\eea
It is immediate that $S_e$ and $S_R$ are invariant under $SL(2,\RR)$, which confirms that the
AdS/CFT motivated regularization procedure of (\ref{3a2}) is also $SL(2,\RR)$-invariant.

\subsection{Extrema of the  entropy solve all constraints}
\label{sec31}

We establish a remarkable equivalence between configurations of the data 
$(x_i, \gamma _i, \kappa_i, \mu_i)$ which satisfy the constraints $\cE_i^{(1)}=\cE_i ^{(2)}=0$
for $i=1,\cdots, N$,
and those which provide extrema of the entropy $S_e$. To formulate this equivalence
precisely, we begin by spelling out the data that are kept fixed, and those that are to be 
varied in the extremization procedure.
 
\sm

The charge vectors $\mu_i$ are subject to overall charge conservation (\ref{2b2}) and 
will be held fixed. The unit vector $\hat \kappa _i$ (satisfying $\hat \kappa_i \cdot \hat \kappa _i=1$)
is taken to be orthogonal to $\mu_i$ and will also be held fixed. Relating the unit vector 
$\hat \kappa _i$ to $\kappa _i, \mu_i$, and  $\gamma_i$ by,
\bea
\label{kappa1}
\kappa _i = \sqrt{2} \, \gamma_i \, \hat \kappa _i
\hskip 1in
\hat \kappa _i \cdot \mu_i =0
\eea
the constraints $\kappa_i ^2=2 \gamma _i^2$ and 
$\kappa_i \cdot \mu_i =0$ of (\ref{2b3}) will automatically hold. 
In terms of the independent variables $x_i, \gamma _i, \hat \kappa _i$, and $\mu_i$
(the last subject to overall charge conservation), the reduced entropy $S_R$ of (\ref{ent1})  
is given by,  
\bea
\label{3b1}
S_R = \sum _{i=1}^N \sum _{j \not= i} \left ( 
{ 2 \gamma_i \gamma_j (1- \hat \kappa _i \cdot \hat \kappa _j) \over (x_i -x_j)^2} 
- 2\sqrt{2} \, { \gamma _i \hat \kappa _i \cdot \mu_j  \over x_i - x_j}
- \half \mu_i \cdot \mu_j \ln { (x_i -x_j)^2 \over \gamma _i \gamma _j} \right )
\eea
We shall now prove that the following variational problem precisely yields 
the   constraint equations $\cE_i^{(1)}=\cE_i ^{(2)}=0$ of (\ref{2b3}) and (\ref{2b4}). 
Keeping the charges $\mu_i$ and the 
unit vectors $\hat \kappa _i$ fixed, and varying freely with respect to $x_i$ produces 
the equations $\cE_i ^{(1)}$. Varying freely with respect to $\gamma_i$ produces 
the equations $\cE_i ^{(2)}$. Indeed, from (\ref{3b1}) one establishes the relations,  
\bea
\label{3b2}
 {\p   S_R \over \p x_i}  = - 2 \cE_i ^{(1)}
 \hskip 1in 
\gamma _i   {\p   S_R \over \p \gamma _i} = \cE_i ^{(2)}
\eea
where the identification $\kappa _i = \sqrt{2} \gamma _i \hat \kappa_i$ has been
used to re-express the result of the variation of $S_R$ in the form on (\ref{2b4}).
 It is remarkable that 
the constraints imposed on the solutions by the BPS conditions and the 
equation of motion can be viewed as conditions which follow from extremizing the 
holographic boundary entropy.  


\section{K\"ahler Structure of Moduli and Calabi's Diastasis}
\setcounter{equation}{0}
\label{sec4}

In this section, we shall exhibit the K\"ahler structure which underlies the moduli space of 
half-BPS solutions with $N$ asymptotic $AdS_3 \times S^3$ regions. We shall also introduce 
Calabi's diastasis function in this setting, and relate it to the scalar fields in Type 4b supergravity. 
In subsequent sections, the entanglement entropy of certain subclasses of these solutions 
will be expressed with the help of the diastasis function.

\subsection{K\"ahler structure of moduli}

In Type 4b supergravity, the scalar field $V$ takes values in 
$SO(5,m)/\left ( SO(5) \times SO(m) \right)$, a Grassmannian which is not generally K\"ahler 
(although it is $m=2$). For half-BPS solutions, however, supersymmetry
requires $V$ to take values in the following submanifold,
\bea
\label{4a1}
\cK \equiv { SO(2,m) \over SO(2) \times SO(m) }
\eea
which is a K\"ahler Grassmannian for any value of $m$. The scalar field provides
a smooth map $V:  \Sigma \to \cK$. Of central interest here are the values of $V$
at the points $x_i$ on $\p \Sigma$, since all the solutions are specified uniquely
by the data  $ x_i, \gamma _i, \kappa _i, \mu_i$ at the $i=1,\cdots, N$ asymptotic 
$AdS_3 \times S^3 $ regions. We have 
learned, either from examination of  the constraints (\ref{2b3}) or from the 
variational solution provided in section \ref{sec31}, that the parameters $ x_i, \gamma_i $ 
are determined in terms of the data $\hat \kappa _1, \cdots , \hat \kappa _N$ and $ \mu_1,
\cdots , \mu_N $,  subject to overall charge conservation (\ref{2b2}) and the conditions 
$\hat \kappa _i^2=1$, $\mu_i^2 >0$, and $\hat \kappa _i \cdot \mu_i=0$.

\sm

For each $i=1,\cdots, N$ the pair $(\hat \kappa _i, \mu_i)$, with $\hat \kappa _i, \mu_i \in \RR^{2+m}$
subject to the conditions $\hat \kappa _i^2=1$, $\mu_i^2 >0$, and $\hat \kappa _i \cdot \mu_i=0$,
projects to a unique point in the K\"ahler manifold $\cK$. This follows from the fact that
the two linearly independent vectors $\hat \kappa _i$ and $\mu_i$ uniquely define a  
2-plane in $\RR^{2+m}$, and thus a unique point in the Grassmannian $\cK$.  
An explicit formula may be obtained for the canonical section $\sigma : \cK \to SO(2,m)$
in terms of the scalar field $V$ by
\bea
\label{4a2}
\sigma _{AB} & = & \eta _{AB} - 2 V^+{}_A V^- {}_B - 2 V^- {}_A V^+{}_B
\no \\
& = & \eta _{AB} - 2 \hat \kappa _A \hat \kappa _B - 2 \hat \mu _A \hat \mu _B
\eea
The canonical section $\sigma$ indeed takes values in $SO(2,m)$, as may be verified with
$\sigma ^t \eta \sigma = \eta$, and is clearly invariant under the action of $SO(2) \times SO(m)$,
so that it is properly a map from the coset $\cK$.
Therefore, any pair $(\hat \kappa_i , \mu_i)$ projects to a unique point in $\cK$. The converse, however,  
does not hold. First because given a value $\sigma_i$, only the 2-plane in which $\hat \kappa_i$
and $\mu_i$ live is determined, but the square $\mu_i^2$ and the angle distinguishing the direction of 
$\mu_i$ from  the direction of $\hat \kappa_i$ are not determined by specifying a point 
$\sigma _i$ in $\cK$. 

\sm

Specifying $\mu^2_i \in \RR^+$ in each asymptotic $AdS_3 \times S^3$ region $i=1,\cdots, N$
amounts to specifying the radius of the $AdS_3$ or equivalently the central charge $c_i$
by the Brown-Henneaux formula. Adding  an angle $\theta \in U(1)$ at each point $i$
completes the extra data into a point in $\cK^+= SO(2,m)/SO(m) \times \RR^+$, so that the full
moduli space is given by,
\bea
\label{4a3}
(\cK^+ _1 \times \cdots \times \cK^+_N) _{\hbox{cc}}
\eea
where the subscript ``cc" stands for enforcing the charge conservation relation of (\ref{2b2}).
This space naturally projects to the K\"ahler manifold $\cK_1 \times \cdots \cK_N$ under the 
map provided by $\sigma$ as a function of the scalar field $V$.

\subsection{Calabi's diastasis in terms of supergravity fields}

The purpose of this section is to compute the K\"ahler potential and evaluate Calabi's 
diastasis function for the K\"ahler coset space $\cK$.
The starting point is the frame field $V$ of the $SO(2)\times SO(m)$ principal bundle over 
the coset $SO(2,m)/SO(2)\times SO(m)$, whose total space is the group $SO(2,m)$.
It may be decomposed as follows,
\bea
\label{4b1}
V = \left ( V^{(\rho,r)} {} _A \right ) 
= \left ( \matrix{V^\rho {}_\alpha & V^\rho {}_a \cr V^r {}_\alpha  & V^r {}_a \cr } \right )
\eea
where $A=(\alpha, a)$ runs over the indices of the defining representation of $SO(2,m)$ with 
$\alpha =1,2$ and $a=6, \cdots , m+5$, while $\rho$ and $r$ run over the indices respectively of the 
defining representations of $SO(2)$ and $SO(m)$ with $\rho=1,2$ and $r=6, \cdots, m+5$.
The group $SO(2,m)$ acts on $V$ by right-multiplication, while $SO(2) \times SO(m)$
acts by left-multiplication.

\subsubsection{The K\"ahler form and metric}

To compute the K\"ahler form $\om_\cK $ of $\cK$, we identify $\om _\cK$ with the $SO(2)$ projection of
the curvature of the right-invariant canonical connection $Q^{12}$ of this  $SO(2)\times SO(m)$-bundle.
Supergravity formulas gives us $Q= - Q^t$ in terms of the scalar field $V$ by the formula, 
\bea
\label{4b2}
d V \, V^{-1} = \left ( \matrix{Q & \sqrt{2} \, P \cr \sqrt{2} \, P^t  & S \cr } \right )
\eea
and provide its curvature $\om_\cK = dQ^{12}$ in terms of $P$ by\footnote{Since $Q$ is a 
connection valued in $SO(2)$, the $Q\wedge Q$ term is absent in the formula for the curvature.}
\bea
\label{4b3}
\om_\cK = dQ^{12}   = 2 \sum _r P^{1r}  \wedge P^{2r}
\eea
In terms of the complex components of the scalar fields $V^{1,2}{}_A$ introduced in  (\ref{2a4}), 
the relevant algebraic relations are given by,
\bea
\label{4b5}
\eta ^{AB} \, V^\pm {}_A V^\mp {}_B & = & 1
\no \\
\eta ^{AB} \, V^\pm {}_A V^\pm {}_B & = & 0
\eea
In terms of the variables $V^\pm {}_A$ the K\"ahler form $\om_\cK$ and the K\"ahler metric $ds^2 _\cK$ become, 
\bea
\label{4b6}
\om _\cK & = & - i \eta ^{AB} \, d V^+{}_A \wedge dV^-{}_B 
\no \\
ds^2 _\cK & = &  \eta ^{AB} \,  dV^+{}_A  \, dV^-{}_B
\eea
By construction, the K\"ahler form and metric are invariant under $SO(2,m)$.

\subsubsection{The K\"ahler potential}

To obtain the K\"ahler potential, it will be convenient to fix the gauge for the $SO(2)$ 
which acts on the indices $\pm$ of $V^\pm {}_A$. This will allow us to express the K\"ahler form, 
metric, and potential in terms of local complex coordinates. To do so in practice, we follow 
\cite{calabi} and choose $V^+{}_1 + i V^+{}_2$ to be real. The remaining complex coordinates 
are introduced as follows,
\bea
\label{4b7}
V^+{}_1 = { w +1 \over 2 \, \mN}
\hskip 0.8in
V^+{}_2 = i { w -1 \over 2 \, \mN}
\hskip 0.8in 
V^+{}_{5+s} = { z^s \over \mN}
\eea
where $s=1, \cdots, m$. The equations of (\ref{4b5}) determine $\mN$ in terms of the other variables, 
and give $w$ as a holomorphic function of the matrix  $Z$ defined by $Z^t = (z^1, \cdots, z^m)$, so that,
\bea
\label{4b8}
2 \, \mN^2 = 1+ \left | Z^t Z \right |^2 - 2 Z^\dagger Z \hskip 1in w = Z^t Z
\eea
The domain which represents the coset $\cK$ in the variable $Z$ corresponds to $\mN >0$
along with the choice $Z^\dagger Z< 1$, and  is referred to as the Lie sphere,
\bea
\label{4b9}
\cK = \left \{ Z \in \CC^m ~ \hbox{such that} ~  Z^\dagger Z < \half + \half \left | Z^t Z \right |^2 < 1 \right \}
\eea
Expressing the K\"ahler form of (\ref{4b6}) in terms of these variables gives, 
\bea
\label{4b10}
\om_\cK =  d \left ( i d \ln \mN - i { (Z^tZ) \, Z^\dagger d \bar Z -  Z^t d \bar Z \over  \mN ^2} \right )
 = - i \p \bar \p \ln \mN^2
\eea
with the help of the standard notations, $d = \p + \bar \p$ where $\p = \sum _s dz^s {\p \over \p z^s}$.
The K\"ahler potential $K$, which is defined by $\om _\cK = i \p \bar \p K$, is given by,
\bea
\label{4b11}
K(Z,\bar Z) = - \ln \Big (1+  | Z^t Z |^2 - 2 Z^\dagger Z \Big )
\eea

\subsubsection{Calabi's diastasis function}

Calabi's diastasis function $\cD (1,2)$ is defined for a pair of points in the K\"ahler 
manifold $\cK$. We shall label these points by their complex coordinates 
$Z_a= (z_a ^1 \cdots z_a ^m)^t$ for $a=1,2$ and set $w_a= Z_a^t Z_a$. 
Calabi's diastasis function is then defined by,
\bea
\label{4c1}
\cD (1,2)  \equiv K(Z_1, \bar Z_1) + K(Z_2, \bar Z_2) - K(Z_1, \bar Z_2)- K(Z_2, \bar Z_1)
\eea
In terms of the coordinates $Z$, and the composite $w = Z^t Z$,  it takes the following form,
\bea
\label{4c2}
\cD (1,2) = 
\ln { \left ( 1+ w_1 \bar w_2  - 2 Z_2^\dagger Z_1 \right  ) 
\left ( 1+ w_2 \bar w_1  - 2 Z_1^\dagger Z_2 \right  )
\over  \left ( 1+ \bar w_1 w_1   - 2 Z_1^\dagger Z_1 \right )
\left ( 1+ \bar w_2 w_2   - 2 Z_2^\dagger Z_2  \right )}
\eea
For $Z_1$ and $Z_2$ near the origin, the diastasis function reduces to 
$\cD (1,2) \approx 2(Z_1 - Z_2)^\dagger (Z_1 - Z_2)$, 
and is proportional to the local Euclidean distance. More generally, it is 
an immediate consequence of the definition of the diastasis function in (\ref{diadef})
that locally for $t_2 \approx t_1$, the diastasis function is always approximated by the 
Euclidean distance. Globally, however, the diastasis function and the Riemannian 
distance between two points behave quite differently, both qualitatively and quantitatively.
Key differences are that the diastasis function is neither always positive, not always 
obeys the triangle inequality.

\subsubsection{Recasting the diastasis function in terms of the scalars $V$}

To recast the diastasis potential in terms of the original supergravity scalars $V$,
we begin by using the expression for the functions $2 \mN_a^2=  1+ |w_a|^2  - 2 Z_a^\dagger Z_a$.
Eliminating the combinations in the denominator of the argument of the logarithm in (\ref{4c2}), 
we find, 
\bea
\label{4c3}
\cD (1,2) = 
\ln \left | { 1 + \bar w_1 w_2 -2  Z_1^\dagger  Z_2 \over 2 \, \mN_1 \, \mN_2} \right |^2
\eea
We may now express Calabi's diastasis in terms of the values $V (x_i)$ and $V(x_j)$ 
of the scalar field $V$ at a pair of points $i,j$, using (\ref{4b7})
and their complex conjugates, and we find, 
\bea
\label{4c4}
\cD (i,j) = 
\ln \left | \eta ^{AB} V(x_i)^+ {}_A V(x_j)^- {}_B \right |^2
\eea
Given the asymptotic values of the scalar field provided in (\ref{asymscal}), 
we obtain an equivalent relation directly in terms of the unit vector 
$\hat \kappa _i, \hat \kappa _j$ and  $\hat \mu _i, \hat \mu _j$, as follows,
\bea
\label{4c5}
\cD (i,j) = \ln \Big ( (\hat \kappa _i \cdot \hat \kappa _j + \hat \mu_i \cdot \hat \mu_j )^2 
+ (\hat \kappa _i \cdot \hat \mu _j - \hat \mu_i \cdot \hat \kappa_j )^2 \Big )
\eea
Note that the formulas for $\cD (i,j)$ are manifestly invariant under $SO(2,m)$.


\section{Entanglement entropy and  diastasis  of  interfaces}
\setcounter{equation}{0}
\label{sec5}

In this section we will solve the constraints and evaluate the entanglement entropy for the 
simplest nontrivial case, namely the $N=2$ interface. In this case the general expression 
for the entanglement entropy (\ref{ent1}) takes the following form, 
\bea
\label{sentint}
S_e &=& {\pi^2\over G_N} \left ( 2{2\gamma_1\gamma_2-\kappa_1 \cdot \kappa_2\over (x_1-x_2)^2 }+ 2{\kappa_1\cdot \mu_2-\kappa_2\cdot \mu_1\over x_1-x_2} - \mu_1\cdot \mu_2 \ln(x_1-x_2)^2
\right .
\no \\ 
&& 
\hskip 0.5in 
\left . 
 +{1\over 2} \mu_1^2  \ln {\mu_1^2  \over 4 \kappa_1^2  }
  +{1\over 2} \mu_2^2  \ln {\mu_2^2 \over 4 \kappa_2^2 }+{1\over 2}(\mu_1^2 +\mu_2^2) \ln   {L^2\over   \ep^2} \right )
\eea
Note that charge conservation equates  $\mu_1^A=-\mu_2^A$, which together with  (\ref{2b4})   implies
\be\label{kmucon1}
\kappa_1\cdot \mu_2= \kappa_2\cdot \mu_1=0.
\ee
The constraint $\cE_i ^{(2)}=0$ of  (\ref{2b4})  now  takes the form,
\be
\label{x12con}
\mu_1^2 (x_1-x_2)^2= 4\gamma_1 \gamma_2- 2 \kappa_1\cdot \kappa_2
\ee
and can be used  to eliminate $x_1-x_2$ from the entaglement entropy  (\ref{sentint}).  
In addition  we use (\ref{kappa1})  and  (\ref{2b4})   to replace $\kappa_i$ by the normalized 
$\hat \kappa_i$ for $i=1,2$ and  (\ref{2c3}) to replace $\mu_1^2= \mu_2^2$ by the central 
charge (which is 
the same on both sides due to charge conservation).  The entanglement entropy becomes,
\be
\label{sentb}
S_e= {c\over 3}  \ln {L \over \ep}  + {c\over 6} \big(1-\ln  2 \big)
+  {c\over 12}  \ln \left({1- \hat \kappa_1\cdot \hat \kappa_2 }\right)^2 
\ee
The first term in (\ref{sentb}) is the universal contribution to the entanglement entropy which 
only  depends on the central charge $c$, the length of the interval $L$, and the UV cutoff 
$\ep$ \cite{Holzhey:1994we,Calabrese:2004eu}. It has the same form whether  or not 
an interface is present and can be removed by considering the difference 
between  the entanglement entropy of pure $AdS_3$  and the interface space-time.   
The second term in (\ref{sentb}) is non-universal and can be eliminated by a moduli independent 
rescaling of the cutoff. The third term in (\ref{sentb}) is universal and present for a  
nontrivial interface.   Hence, it may be  
identified with  the $\mg$-function  of the interface  \cite{Affleck:1991tk},
\bea
\label{5e1}
\ln  \mg = {c\over 12} \, \ln  \left( { 1-\hat \kappa _1 \cdot \hat \kappa _2 } \right)^2 
\eea
We can relate  the $\mg$-function  to the geometric diastasis function by 
using (\ref{4c5}), and the fact that $\hat \mu_1 \cdot \hat \mu_2=-1$, so that we find 
the following general expression for the $\mg$-function in terms of the central charge
and the diastasis function of the interface,
\bea
\label{dia6}
2 \ln \mg = {c\over 6}\, \cD(1,2)
\eea
The extra  factor $c/6$ in front of the geometric diastasis function  in (\ref{dia6}) 
compared to (\ref{diarela}) has the following explanation. The underlying CFT of the 
Type 4b $AdS_3\times S^3$ vacua is given by a symmetric product of 
${\cal M}^N/S_N$ where  ${\cal M}=T^4$ for the $m=5$ case and ${\cal M}=K_3$ 
for $m=21$. Since the  CFT  corresponding to a  single ${\cal M}$ 
target space has central charge   $c=6$ this implies that the number of copies of in the 
symmetric product  $N=c/6$.  Note that in a symmetric product all copies of the 
underlying $\cM$  CFT are at the same point in the moduli space. Consequently  
the diastasis function for the ${\cal M}^N/S_N$  symmetric product CFT is given by 
$N$ times the diastasis function of the underlying CFT with target space ${\cal M}$. 
The example of \cite{Bachas:2013nxa} comprises a target space which is a single 
copy of $T^2$ and hence (\ref{diarela})  holds without any additional factor.

\sm

To summarize, we have demonstrated the holographic version of the relation (\ref{diarela}) 
between the $\mg$-function and the Calabi's diastasis function, which was first discovered 
for the dual interface CFTs  in \cite{Bachas:2013nxa}.


\section{Entanglement entropy and diastasis for 3-junctions}
\setcounter{equation}{0}
\label{sec6}

The goal of this section is two investigate whether the entanglement entropy (\ref{ent1}) 
for the $3$-junction can be related to the diastasis function. For the case  $N=3$  
the constraint equations  $\cE_i ^{(1)}=0$  in (\ref{2b3}) imply the $\cE_i ^{(2)}=0$ constraint. 
Despite this simplification, the $N=3$ case is still considerably more difficult than  the $N=2$ 
case of the interface, treated in the previous section, due in part to the fact that for a 3-junction, 
charge conservation does not force the charges $ \mu^A$ to be parallel.  
The constraint equations form a non-linear system whose complete solution 
appears to require solving a quintic equation of general type. 

\sm

To make progress, we recast the entropy and the constraints in terms of 
manifestly $SL(2,\RR)$-invariant variables, which are defined as follows,
\bea
y_i = { \gamma _i (x_j-x_k) \over (x_i-x_j) (x_i-x_k)}
\hskip 1in 
\Delta _{ij} = \hat \kappa _i \cdot \hat \kappa _j -1
\eea
where $(i,j,k)$  in the first equation is a cyclic permutation of $(1,2,3)$.
In these variables, the constraint equations become quadrics, 
\bea
\label{constrain}
0 = \mu_i^2 + \sqrt{2} \hat \kappa _i \cdot (\mu_j - \mu_k ) y_i
+ 4 y_i y_j \Delta _{ij}  + 4 y_i y_k \Delta _{ik} 
\eea 
Successively eliminating two of the three $y_i$  produces a  polynomial equation
in the third variable of degree 5, which does not lead to algebraic solutions. 
In terms of these variables, the entropy $S_R$ given in (\ref{3b1}) takes the form,
\bea 
\label{6a4}
S_R&=&  - 4  y_1 y_2 \Delta _{12} - 4 y_2 y_3 \Delta _{23}  - 4  y_3 y_1  \Delta _{31}
+{1\over 2}\sum_{i=1}^3 \mu_i^2 \ln   {1 \over y_i^2}
\no \\
&&
+  \sqrt{2} \hat \kappa_1 \cdot (\mu_3-\mu_2)  y_1
+ \sqrt{2} \hat \kappa_2 \cdot (\mu_1-\mu_3)  y_2
+ \sqrt{2} \hat \kappa_3 \cdot (\mu_2 -\mu_1)  y_3
\eea
Note that the variables $y_i$ now encompass the free $SL(2,\RR)$-invariant combinations of the 
variables $x_i, \gamma _i$, so that extremization in $y_i$ indeed reproduces the constraint
equations (\ref{constrain}).

\subsection{Solving for constrained charges}

Although it does not appear possible to solve in simple terms for the 3-junction entropy in all generality,
it is nonetheless possible to solve for a subclass of physically interesting charge arrangements.
Under the assumption that the vector space spanned by the vectors $\kappa_i$ is orthogonal to the vector space 
spanned by the vectors $\mu_i$, the entropy may be obtained in explicit form, and in fact exhibits
remarkable properties. This restricted case includes the physically important special situation 
where all 3-form charges $\mu_i$ are parallel to one another. 

\sm

Concretely, the above orthogonality conditions are expressed by,
\bea
\hat \kappa _i \cdot \mu _j =0
\eea
for all $i,j=1,2,3$.  The expression for the reduced entropy of (\ref{6a4}) is given by, 
\bea
S_R = - 4  y_1 y_2 \Delta _{12} - 4 y_2 y_3 \Delta _{23}  - 4  y_3 y_1 \Delta _{31}
+\half \sum _{n=1}^3  \mu_n^2 \ln {1 \over y_n^2}
\eea
while the constraints of (\ref{constrain}) reduce to the equations, 
\bea
0 = \mu_i ^2 + 4  y_i y_j \Delta _{ij} + 4  y_i y_k \Delta _{ik}
\eea
where $(i,j,k)$ is a cyclic permutation of $(1,2,3)$.
Solving for  $y_i y_j$ with $i \not= j$ one finds,
\bea
y_i y_j  =  { \mu_i \cdot \mu_j  \over 4 \Delta_{ij} } 
\eea
Substituting into the entropy, we find,
\bea
S_e&=& 
{1\over 24} ({c_1+c_2-c_3}) \ln \left ( { \Delta_{12} \over \hat \mu_1 \cdot \hat \mu_2} \right ) ^2
+{1\over 24} ({c_2+c_3-c_1}) \ln \left ( { \Delta_{23} \over \hat \mu_2 \cdot \hat \mu_3} \right ) ^2
\no \\ && 
+{1\over 24} ({c_3+c_1-c_2}) \ln \left ( { \Delta_{31} \over \hat \mu_3 \cdot \hat \mu_1} \right ) ^2
+ {1\over 12} (c_1+c_2+c_3) \Big (1 +  \ln {  L^2 \over 2 \ep^2} \Big ) 
\eea
Therefore the entanglement of the 3-junction with the restricted charge assignments 
$\hat \kappa _ i \cdot \mu_j=0$ may be expressed in terms of the diastasis function for 
pairs, since we have,  
\bea
\Delta _{ij} = - 1 - \hat \mu _i \cdot \hat \mu_j 
+ e^{\cD(i,j)/2} \, \hbox{sign} ( \hat \kappa _i \cdot \hat \kappa _j + \hat \mu_i \cdot \hat \mu_j)
\eea
A thorough discussion of the signs involved will be presented in the next section.
Suffice it here to add the following explicit expression which suffice to evaluate the case
where all flux charges $\mu_i$ are parallel to one another,
\bea
\hat \mu _i \cdot \hat \mu_j = -1 & \hskip 1in &
\ln \Delta _{ij}^2 = \cD(i,j)
\no \\
\hat \mu _i \cdot \hat \mu_j = +1 & \hskip 1in &
\ln \Delta _{ij}^2 = \ln \left ( 2
- e^{\cD(i,j)/2} \, \hbox{sign} ( \hat \kappa _i \cdot \hat \kappa _j + 1) \right )^2
\eea
The first of these relations was used for the interface entropy.


\section{Entropy and Calabi's diastasis for $N$-junctions}
\setcounter{equation}{0}
\label{sec7}

For general $N>3$ junctions the system of constraint equations lends itself even less than 
for $N=3$ to a complete solution, as the  constraints $\cE_i ^{(1)}=0$  in (\ref{2b3}) now 
impose further non-trivial relations. Nonetheless, the system may be well-understood in terms
of Calabi's diastasis function for large physically relevant classes of data $\hat \kappa _i, \mu_i$.
Basically, the system lends itself to solution better when the dimension of the vector
space spanned by the charge vectors $\mu_i$ is smaller. We begin by solving the case when the 
dimension is 1, and then produce extensions to low dimensions.

\subsection{Parallel charge vectors $\mu_i$}

One class of data $\hat \kappa _i, \mu_i$ allows for complete solution in terms of Calabi's 
diastasis function, namely when the charge vectors $\mu_i$ are parallel  to one another 
for all $i=1,\cdots, N$, 
\bea
\label{7a0}
\mu_i = \alpha _i \, \hat \mu_1 
\hskip 1in 
\sum _{i=1}^N \alpha _i=0
\hskip 1in 
c_i = { 3 \pi^2 \, \alpha _i^2 \over 2 G_N}
\eea
Above we have indicated the overall charge conservation relation on the coefficients $\alpha _i$, 
as well as the relation resulting from (\ref{2c3}) between the real coefficient 
$\alpha_i$  and the central charge $c_i$ of the $i$-th asymptotic region $AdS_3 \times S^3$. 
Henceforth, we shall replace the data of the positive central charges $c_i$ by those of the coefficients 
$\alpha_i$. Clearly, $c_i$ determines $\alpha_i$ up to its sign, which will be important, and which will be denoted by $\hat \alpha _i =  \hbox{sign} (\alpha _i)$. 

\sm

As a result of the assumption that  $\mu_i$ are all parallel to one another, the orthogonality 
relations $\hat \kappa _i \cdot \mu_i=0$ imply the following orthogonality relations
valid for all $i,j=1,\cdots, N$, 
\bea
\label{7a1}
\hat \kappa _i \cdot \mu _j =0
\eea
The reduced entropy of (\ref{3b1}) simplifies accordingly, and is given by,
\bea
\label{7a2}
S_R = \sum _{i=1}^N \sum _{j \not= i} \left ( 
{ 2 \gamma_i \gamma_j (1- \hat \kappa _i \cdot \hat \kappa _j) \over (x_i -x_j)^2} 
- \half \alpha_i \alpha_j \ln { (x_i -x_j)^2 \over \gamma _i \gamma _j} \right )
\eea
The parameters $\alpha_i$ are fixed by the central charges up to signs. 
We seek to eliminate the dependence on the data $\hat \kappa _i $ 
in favor of the diastasis function. The variables $x_i$ and $\gamma _i$ are to
be determined by extremizing the entropy for given $\alpha _i$ and $\hat \kappa _i$,
following section \ref{sec31}.

\sm

Calabi's diastasis function, evaluated between pairs labeled by $i,j$ as in (\ref{4c5}),
then takes on a considerably simplified form under the assumption of (\ref{7a1}), and we have,
\bea
\label{7a3}
\cD (i,j) =  \ln \left (  \hat \kappa _i \cdot \hat \kappa _j + \hat \alpha _i \, \hat \alpha _j \right ) ^2
\eea
Inverting this relation to obtain $\hat \kappa _i \cdot \hat \kappa _j$ in terms of $\cD(i,j)$ and 
$\hat \alpha _i \hat \alpha _j$ requires care with sign issues.
While the manifold of unit vectors $\hat \kappa_i $ in $\RR^{2+m}$ with signature $(++- \cdots -)$
is connected, its submanifold of unit vectors orthogonal to a unit vector $\hat \mu_1$ of positive square 
is disconnected. Its two connected components may be distinguished by a sign $\hat s_i$, 
obtained as follows.
Upon making an $SO(2,m)$ rotation, we may choose a canonical direction for the vector $\hat \mu_1$,
and combine the first relation of (\ref{7a0}) with (\ref{7a1}), to parametrize $\kappa _i$ as follows,
\bea
\label{7a4}
\hat \mu _1= (1,0; {\bf 0} )
\hskip 1in 
\hat \kappa _i = \left (0,  s_i  ; ~ \bs_i  \right ) 
\hskip 1in 
s_i^2 - \bs_i^2 =1
\eea
where $ \hat s_i= \hbox{sign}(s_i)$ while $\bs_i$ is an arbitrary vector in $\RR^m$.
From this parametrization, the following inequality follows right away,
\bea
\label{7a5}
\hat s_i \hat s_j \, \hat \kappa _i \cdot \hat \kappa _j \geq 1
\eea
with equality only when $\hat s_i \, \bs_i = \hat s_j \, \bs_j$. Recasting the diastasis function in
the form, 
\bea
\label{7a6}
\cD (i,j) =  \ln \Big (  \hat s_i \hat s_j \hat \kappa _i \cdot \hat \kappa _j 
+ \hat s_i \, \hat s_j \, \hat \alpha _i \, \hat \alpha _j \Big ) ^2
\eea
it is now straightforward and unambiguous to solve for the combination 
$\hat s_i \hat s_j \, \hat \kappa _i \cdot \hat \kappa _j $ which is always positive by  (\ref{7a5}).
Extracting $\hat \kappa _i \cdot \hat \kappa _j$ from this result, we find, 
\bea
\label{7a7}
\hat \kappa _i \cdot \hat \kappa _j = - \hat \alpha _i \hat \alpha _j + \hat s_i \hat s_j \, e^{\cD (i,j)/2}
\eea
Substituting this result into (\ref{7a2}) gives the desired expression for the reduced
entropy in terms of Calabi's diastasis function, 
\bea
\label{7a8}
S_R = \sum _{i=1}^N \sum _{j \not= i} \left ( 
{ 2 \gamma_i \gamma_j \left (1+ \hat \alpha _i \hat \alpha _j - \hat s_i \hat s_j \, 
e^{\cD (i,j)/2} \right ) \over (x_i -x_j)^2} 
- \half \alpha_i \alpha_j \ln { (x_i -x_j)^2 \over \gamma _i \gamma _j} \right )
\eea
The solutions for $N=2$ and $N=3$ may be derived explicitly, and were given in the 
preceding sections. For $N\geq 4$, the constraint equations at present appear prohibitive.

\subsection{Two-dimensional space of charge vectors $\mu_i$}

We shall now proceed to the case where the space of charge vectors is 2-dimensional.
To simplify the discussion, we limit attention to the case where this space is 
orthogonal to the vectors $\hat \kappa _i$ for all $i=1,\cdots, N$, or equivalently,
\bea
\label{7b1}
\hat \kappa _i \cdot \mu_j=0
\eea
for all $i,j=1,\cdots, N$. Using an $SO(2,m)$ rotation to set $\hat \mu_1$ to the direction 
given in (\ref{7a4}), we see that the orthogonality of $\hat \kappa _i$ with $\hat \mu_1$
forces $\hat \kappa _i$ to take the form given in (\ref{7a4}) for all $i=1,\cdots, N$. Using 
the residual $SO(1,m)$ group which leaves $\hat \mu_1$ invariant, we may choose 
any linearly independent unit charge $\hat \mu_2$ to take the form, 
\bea
\label{7b2}
\hat \mu_2= (\alpha ,\gamma ; \beta , 0 \cdots 0)
\hskip 1in \alpha ^2 + \gamma ^2-\beta ^2=1
\eea
We can have either $\beta=0$ when the vector restricted to $SO(1,m)$ has positive square, or $\gamma=0$
when it has negative square. The case $\gamma \not=0, \beta =0$ is ruled out by orthogonality to 
$\hat \kappa _i$, so that only the case $\gamma =0, \beta \not= 0$ remains, and the first component 
of $\bs_i$ must vanish. In summary, we have for all unit charge vectors,
\bea
\label{7b3}
\hat \mu_j & = & (\alpha _j , 0; \beta _j , 0 \cdots , 0) 
\no \\
\hat \kappa _i & = & (0, s_i ; 0, \bs_i) 
\eea
where $\alpha _j^2 - \beta _j^2 =1$, $\beta _1=0$, and $\bs_i$ is an arbitrary  vector in 
$\RR^{m-1}$ with $s_i^2 - \bs_i^2 =1$.  

\sm

The diastasis function on pairs is now given as follows,
\bea
\label{7b4}
\cD (i,j) = \ln \left ( \hat \kappa _i \cdot \hat \kappa _j + \alpha _i \alpha _j - \beta _i \beta _j \right )^2
\eea
Using care with the signs $\hat s_i$, we may invert this relation to find, 
\bea
\label{7b5}
\hat \kappa _i \cdot \hat \kappa _j = - \alpha _i \alpha _j + \beta _i \beta _j + \hat s_i \hat s_j \, e^{\cD(i,j)/2}
\eea
The reduced entropy now takes the form,
\bea
\label{7b6}
S_R & = & \sum _{i=1}^N \sum _{j \not= i} \left  ( 
{ 2 \gamma_i \gamma_j \over (x_i -x_j)^2} 
\Big \{ 1+  \alpha _i  \alpha _j -\beta _i \beta _j - \hat s_i \hat s_j \,  e^{\cD (i,j)/2} \Big \} \right .
\no \\ && \hskip 1in \left .
- \half \sqrt{\mu_i^2 \mu_j^2} \, (\alpha_i \alpha_j - \beta _i \beta _j) \,
\ln { (x_i -x_j)^2 \over \gamma _i \gamma _j} \right )
\eea
The physical value of the junction entropy is then obtained by the solution to the variational 
problem, exhibited  in section \ref{sec31}, which gives $x_i, \gamma _i$ in terms of the
data $\hat \kappa _i, \mu_i$.

\sm

To summarize, we see that the $N$-junction entropy is  governed 
by Calabi's diastasis function, along with the central charges $c_i = \mu_i^2$, but that 
a further dependence on the ``angles" associated with the hyperbolic
unit vectors $(\alpha _i, \beta _i)$ with $\alpha _i^2 - \beta _i^2=1$ necessarily enters
as well.


\section{Entropy of a non-supersymmetric interface}
\setcounter{equation}{0}
\label{sec8}

The purpose of this section is to show that  the interface entropy of a  non-supersymmetric 
Janus solution is not given by the diastasis function but instead determined by the geodesic 
distance on the moduli space between the theories on the two sides of the interface.  
We include this calculation  for two reasons.  First,  it provides an illustration that diastasis 
and supersymmetry are intimately linked (a point already made in \cite{Bachas:2013nxa}). 
Second,  to the best of our knowledge, the calculation has not appeared earlier in the literature, 
and deserves  attention in its own right.

\sm

The model we consider here is gravity in three-dimensional  space-time 
with local coordinates $x^\mu$ and space-time 
metric\footnote{In this section, $\mu=0,1,2$, while $i,j=1,\cdots, N=\hbox{dim}(M)$, 
and repeated indices are to be summed over.} $g_{\mu \nu} dx^\mu dx^\nu$ which is
coupled to a nonlinear sigma model on a manifold $M$ with Riemannian internal metric. 
In terms of real local coordinates $\phi^i$ on $M$, the internal metric 
is given by $G_{ij}(\phi)d\phi^i d \phi^j$ and is considered fixed, and the action is a 
functional of the space-time metric $g_{\mu \nu}(x)$ 
and the fields $\phi ^i (x)$, given by,
\be
I [g, \phi] = \int d^3 x \sqrt{g} \Big(R-2\Lambda -{1\over 2} G_{ij}(\phi) \partial_\mu \phi^i \partial^\mu \phi^j\Big)
\ee
where $\Lambda$ is the cosmological constant, and $g= - \det (g_{\mu \nu})$.
Einstein's equations are given by, 
\bea
R_{\mu\nu}+ \Lambda g_{\mu\nu}-{1\over 2} G_{ij}(\phi) \partial_\mu \phi^i \partial_\nu \phi^j = 0
\label{graveqa}
\eea
while the scalar field equations are given by, 
\bea
{1\over 2}\Big({\partial \over \partial \phi_k} G_{ij}(\phi) \Big) 
\partial_\mu \phi^i \partial^\mu \phi^j
-{1\over \sqrt{g}} \partial_\mu\Big(G_{ik} \sqrt{g} \, \partial^\mu \phi^k\Big) = 0
\eea
We shall set $\Lambda=1$ and use the following Janus Ansatz in which the space-time 
metric is parameterized by an $AdS_2$ slicing and the scalars only depend on the slicing 
coordinate $y$, 
\be
\label{metra}
ds^2= dy ^2 + f(y) {dz^2-dt^2\over z^2}
\hskip 1in  
\phi^i=\phi^i(y)
\ee
With this Ansatz the $tt$ and $zz$ components of  (\ref{graveqa}) reduce to,
\be
2 -4 f + {d^2 f \over dy ^2} =0
\ee
which is solved by the following  family of  functions dependent on a real parameter $\gamma$,
\be
f(y)= {1\over2 } \left ( 1+ \sqrt{1-2\gamma^2} \cosh 2y \right )
\ee
The $AdS_3$ vacuum solution corresponds to setting $\gamma=0$,
while $2 \gamma^2=1$ corresponds to $\RR \times AdS_2$.  More generally,
regular real solutions correspond to $2 \gamma ^2 \leq 1$.
The $yy$ component of the gravitational equation, and the scalar field 
equation may be similarly derived. Actually, it is illuminating to change 
variables from $y$ to $\lambda$ using the relation, 
\be
f(y) {d\over dy}  =  {d\over d \lambda} 
\ee
The remaining equations then take the form, 
\bea
\label{constra}
G_{ij}(\phi) \, \dot \phi^i \dot \phi^j = 2 \gamma^2
\hskip 1in 
\ddot \phi^i + \Gamma^{i}_{\; jk} \, \dot \phi^j \dot \phi^k = 0
\eea
where the dot stands for differentiation with respect to $\lambda$, 
the Levi-Civita connection of the internal metric $G_{ij}$ is denoted by $\Gamma ^i _{jk}$,
and we continue to use the notation $\phi^i$ now for functions of the coordinate $\lambda$
instead of $y$.

\subsection{Interface entropy}

In this section calculate the  interface entropy   the non-supersymmetric Janus solution 
presented in the previous section. Following \cite{Azeyanagi:2007qj,Chiodaroli:2010ur}, 
the interface entropy (or equivalently the $\mg$-function) can be related to the 
entanglement entropy of a region  which  encloses the interface symmetrically,
\be
\label{entangent}
S_e={c\over 3} \ln {L\over \ep} + \ln \mg
\ee
The holographic calculation of the interface entropy was performed in \cite{Azeyanagi:2007qj}, 
and results in, 
\be
\label{metrb}
\ln \mg = - {c\over 6} \ln \sqrt{1-2\gamma^2}
\ee
Note that we have $g\to 1$ as $\gamma\to 0$, as this limit should indeed correspond to the 
the absence of the  interface, and thus the vanishing of the interface entropy. Also note that 
reality  of  the entanglement entropy in (\ref{metrb}) impose the same  bound on $\gamma$
which had been imposed by the reality of the metric itself, namely $2 \gamma ^2\leq 1$.

\sm

 The constraint equation in (\ref{constra}) relates the geodesic distance $\Delta \ell$ 
 in the internal space of the  non-linear sigma model to the deformation parameter $\gamma$,
\be
\Delta \ell = \int d\lambda \sqrt{G_{ij}(\phi) \, \dot \phi^i \dot \phi^j} = \sqrt{2} \gamma \int d\lambda 
\ee
The geodesic distance is evaluated as follows, 
\bea
\Delta \ell 
= \sqrt{2} \gamma  \int_{-\infty}^{+\infty} d\mu {1\over f(\mu)} 
=  4 \, {\rm tanh}^{-1} \left( {1-\sqrt{1-2\gamma^2}\over \sqrt{2} \gamma}\right)
\eea
Solving for $\gamma$ in terms of $\Delta \ell$, one obtains,
\be
\label{delgamrel}
\gamma= {1\over \sqrt{2}} \tanh\left({\Delta \ell  \over 2}\right)
\ee
Hence we obtain  the interface entropy in terms of the geodesic distance $\Delta \ell$,
\be
2\ln \mg = {c\over 6} \, \ln \,  \cosh \left ({ \Delta \ell \over 2} \right )^2 
\ee
Note that the diastasis function coincides with the geodesic distance in the limit of 
infinitesimal separation \cite{Bachas:2013nxa} but differs for finite separations.   
Consequently, since the geodesic distance is different from the diastasis function 
the condition that the interface is BPS is essential in connecting $\ln \mg$ to the diastasis function.

\sm

Equation (\ref{delgamrel}) implies that an expansion in small separation is equivalent 
to an expansion in small $\gamma$.  The agreement of the diastasis and geodesic 
distance is consistent with  \cite{Azeyanagi:2007qj} where it was found that the BPS 
and non-BPS interface $\mg$ functions agree at first order in an expansion in  $\gamma$.


\section{Summary and Discussion}
\setcounter{equation}{0}
\label{sec9}

The main results of this paper may be summarized as follows.

\sm

First, we have proven that the equivalence  of interface entropy and Calabi's diastasis function, 
which was derived for BPS interfaces in certain two dimensional CFTs in \cite{Bachas:2013nxa}, 
continues to hold holographically for BPS interface solutions in six-dimensional Type 4b supergravity. 
Key to this equivalence is the fact that the moduli space of half-BPS interface solutions in 
Type 4b supergravity has an underlying K\"ahler manifold structure, which makes the 
appearance of Calabi's diastasis function possible.

\sm

Second, we have extended the application of entanglement entropy to the case of a
junction of $N\geq 3$ CFTs, where we have defined and carefully regularized an associated 
junction entropy. Using the holographic realization of these $N$-junctions in terms of 
Type 4b supergravity solutions, we have identified the moduli of the solutions, 
exhibited their underlying K\"ahler structure, and produced a variational formula 
for the evaluation of this junction entropy. For special arrangements of the flux charges,
including when all the flux charges are parallel to one another, we have shown that 
the junction entropy may be represented as a sum of Calabi's diastasis functions evaluated between
the data associated with pairs of asymptotic $AdS_3 \times S^3$ regions.

\sm

Third, we have shown that the interface entropy of a non-supersymmetric 
Janus solution to a 3-dimensional gravity-non-linear-sigma-model for a general 
internal Riemannian manifold $M$ is given in terms of  the geodesic distance on $M$,
and not in terms of any diastasis function. Unless $M$ is K\"ahler the 
diastasis function would not even exist. We interpret the results of this calculation as 
lending support to the assertion that  the entropy-diastasis equivalence  is intimately 
connected with supersymmetry.

\sm

The results in this paper leave open several interesting questions and avenues for future research, 
of which we list  the following.

\sm

It would be instructive to find a way to relax the orthogonality condition  (\ref{7b1}) on the moduli 
of the $N$-junction solution, and to obtain the junction entropy for general charge assignments. 
Even for the 3-junction the general case appears considerably more complicated to solve algebraically, 
as it involves solving a quintic equation.  
It may be that a better parameterization, possibly along the lines of the light cone like 
variables used in \cite{Chiodaroli:2011nr}, might help to solve the general case. 
Since the orthogonality condition were only imposed as a means of making the 
constraint equations solvable,  it would   be interesting to determine whether this 
condition  has any  physical meaning on the CFT side.

\sm

In Calabi's original paper \cite{calabi} the diastasis function is regarded as a potential.
Specifically, the function $\cD(1,2)$ is interpreted as the ``potential" at point $2$ in the 
presence of a ``source" at point $1$. A natural question emerging from this work is whether 
the $N$-junction entropy can be usefully interpreted as a potential in the presence of $N-1$ 
sources as well. One encouraging piece of supporting evidence is the fact that the 
constraints of (\ref{2b3}) may be obtained equivalently from the variation of the entanglement 
entropy  which, in turn,  may be viewed as a zero force condition  for a potential.

\sm

It would also be interesting to study the junction entropy on the 
CFT side along the lines of the work in \cite{Bachas:2013nxa} for interfaces. 
For example, we have already found that the holographic entanglement entropy for a 3-junction 
decomposes into a sum of diastasis functions  between pairs, weighed by 
combinations of the three central charges. It would be valuable to 
determine whether such a structure can arise for BPS junctions on the CFT side as well.

\bigskip \bigskip

\noindent{\Large \bf Acknowledgements}

\medskip

We  acknowledge useful conversations with Constantin Bachas, Michael Douglas, and Simon Gentle.
This work was supported in part by National Science Foundation grants PHY-13-13986 and PHY-11-25915.
One of us (ED) thanks the Kavli Institute for Theoretical Physics at the University of California, 
Santa Barbara for their hospitality and the Simons Foundation for their financial support while part
of this work was being carried out. 


\appendix

\section{Regularization}
\setcounter{equation}{0}
\label{appb}

In this appendix we present a careful holographic UV regularization and 
exhibit how the cutoff is imposed 
on the asymptotically $AdS_3 \times S^3$ regions  for the $N$-junction solutions.

\subsection{Minimal area surface}

In this subsection we adapt an argument  \cite{Jensen:2013lxa,Estes:2014hka} concerning the 
minimal area hyper-surface for $AdS_2\times S^2$ fibrations, where the metric is given by (\ref{2a1}). 
The metric of the $AdS_2$ factor is given by the unit radius Poincare patch metric
\be
ds_{AdS_2}^2= {dz^2- dt^2\over z^2}
\ee
The static minimal surface which is used to calculate the holographic entanglement entropy 
is independent of $t$, spans the sphere $S^2$ and  extends over the  Riemann surface $\Sigma$. 
Since the $SO(3)$ isometry of the two sphere is unbroken in the solution, the embedding is independent 
of the coordinates of $S^2$ and only depends on the local coordinates $w,\bar w$ of $\Sigma$,
so that the embedding is completely specified by a single real-valued function $z(w,\bar w)$ of $\Sigma$. 
The entanglement entropy $S_e$ is  given by the area of the  four-dimensional hyper-surface  
which minimizes the  area with respect to the metric 
$G_{\rm ind}$ induced on $S^2 \times \Sigma$,
\bea
\label{entareaf}
S_e & =&{1\over 4G_{N}} \int_{S^2 \times \Sigma} d^4\xi  \sqrt{G_{\rm ind}}
\nonumber\\
&=& {\hbox{Vol}(S_2) \over 4G_{N}}   \int |d w|^2 \, F_-^2 \, \sqrt{1+ {H^2 \over F_-^2  z^2} \left( {\partial z \over\partial w} {\partial z\over \partial{\bar w}}\right)}
\eea
The square root on the second line above is manifestly bounded from below by 1,
and this lower bound is uniquely attainted when the function $z$ is constant. 
Thus,  $z=z_0$ constant is a solution and gives the absolute minimum for the area. 
The entanglement entropy on the hyper-surface of minimal area thus takes the form,
\be
\label{sentapp}
S_e= {\hbox{Vol}(S_2) \over 4G_{N}}   \int |dw|^2 \; F_-^2 
\ee
The integration has divergent contributions coming from the asymptotic $AdS_3$ regions near $w=x_n$, $n=1,2,\cdots N$, which we shall regularize in the sequel.

\subsection{Poincar\'e and $AdS_2$ slicing}

In order to make a connection with the field theory  result and extract the boundary entropy we have to carefully identify the cutoff $\ep_i$ employed in the regularization of the entanglement entropy  (\ref{sentapp}) with the UV cutoff in the field theory. The UV cutoff is defined by mapping the asymptotic metric near $w=x_i$  into a Fefferman-Graham coordinate system.

We start with an illustrative example mapping the three dimensional $AdS$ metric 
in Poincar\'e coordinates $(u, \eta, t)$, and metric,
\bea
\label{poincare}
ds^2 = R^2 \, { d u ^2 + d \eta ^2 - dt^2 \over u^2}
\eea
to   new set of coordinates $(x,z,t)$ 
\bea\label{changev}
u = { z \over \cosh x} 
\hskip 1in 
\eta = z \tanh x
\eea
In terms of these new coordinates $(x,z,t)$, the metric is the $AdS_2$ slicing of $AdS_3$,
\bea\label{ads2sli}
ds^2 = R^2 \left ( dx^2 + \cosh^2 x \, { dz^2 - dt^2 \over z^2} \right )
\eea
The Poincar\'e coordinates (\ref{poincare}) are already in Fefferman-Graham form and  
the boundary is reached by taking $u\to 0$.   The map (\ref{changev}) shows that the 
boundary $AdS_2$ slicing (\ref{ads2sli}) has three components:  the $AdS_2$ boundary 
$z\to 0$ which we identify with the interface and the  two asymptotic regions $x\to \pm \infty$. 
For the latter regions and $z=z_0$ finite we can relate the cutoff in the Poincare coordinates 
$u=\ep$ with the cutoff in $x$ by (\ref{changev}).

\be\label{cutoffa}
e^{-|x|_{\ep}}  = {\ep\over z_0} 
\ee
For the solutions which are discussed in the present paper the situation is more 
complicated in several ways:
\begin{enumerate}
\itemsep=0.0in
\item The  three coordinates  $x,z,t$ are accompanied by   three  additional  coordinates 
parameterizing the two sphere and an additional (angular) slicing coordinate.
\item  The metric (\ref{ads2sli}) has two asymptotic regions $x\to \pm \infty$ and describes 
a configuration where two half spaces are glued together at an interface.  For general $N>2$ 
our solutions  describe junctions where $N$ half spaces are glued together.
\item  The map (\ref{changev}) covers the complete  Poincar\'e patch. Even for interface solutions 
a globally defined map is not known and the map has to  be defined in patches.  
\end{enumerate}
In the following we shall address some of these issues with the primary goal to generalize 
the identification (\ref{cutoffa}) to our solutions.

\subsection{Regularization}

To calculate the form of the metric near the $n$-th asymptotic region $w=x_i$, it is 
convenient to introduce the coordinate $w= x_i + e^{-x+i \theta}$,  where the 
asymptotic $AdS$ region is reached by taking $x\to \infty $   and $\theta \in [0,\pi]$.
 The metric takes the following form
 \be
 \label{adsmet1}
 ds^2= \rho^2 e^{-2x} (dx^2+ d\theta^2)+f_1^2 {dz^2-dt^2\over z^2}+ f_2^2 ds_{S^2}^2
 \ee
 the asymptotic behavior  of the metric near $w=x_n$ can be extracted from (\ref{2a1}) and (\ref{2a2})
   \bea
   \label{asymmeta}
\rho^2 &\sim &R_i^2\;   e^{2x} \; \Big(  1+ e^{-x} \rho^{(1)}(\theta) + e^{-2x} \rho^{(2)}(\theta)+\cdots\Big) 
\nonumber\\
f_2^2 &\sim&   R_i^2  \;  \sin^2 \theta   \Big( 1+ e^{-x} f_2^{(1)}(\theta) + e^{-2x} f_2^{(2)}(\theta)+\cdots\Big) 
\nonumber\\
f_1^2 &\sim &R_i^2\;   {A_i^2 \over 4} e^{2x}  \Big( 1 + e^{-x} f_1^{(1)}(\theta)+ e^{-2x} f_1^{(2)}(\theta)+\cdots    \Big)
 \eea
 
 \begin{figure} [htb]
\centering
\includegraphics[scale=.7]{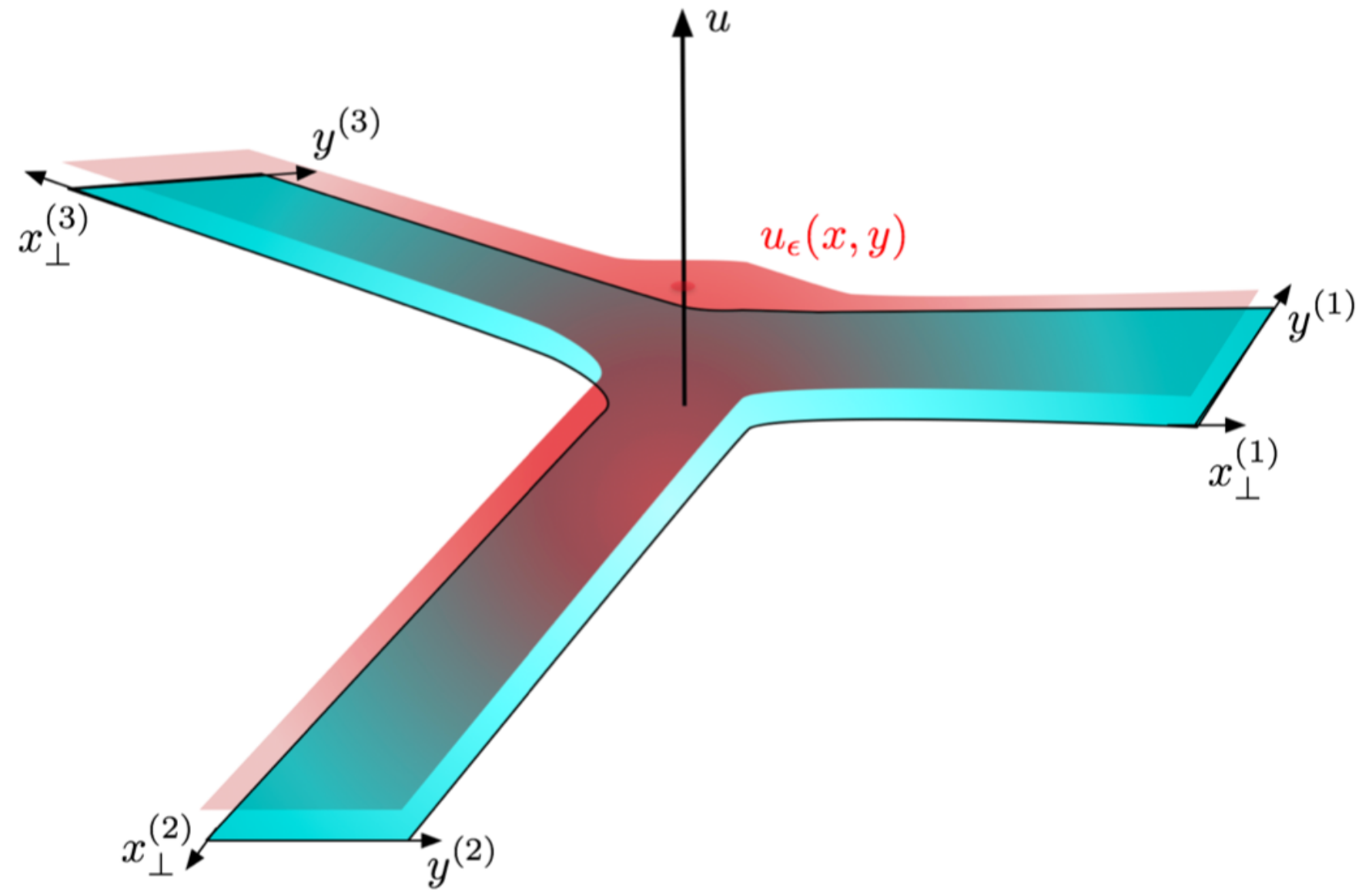}
\caption{ Fefferman-Graham cutoff surface $u=u_{\epsilon}(x,y)$ for $N=3$ junction.}
\label{fig3}
\end{figure}

Here the dots  in the brackets denote terms which fall off faster than $e^{-2x}$ in the limit $x\to \infty$. 
The $AdS$ radius and the constant $A_i>0$ are expressed in terms of moduli associated 
with the $i$-th asymptotic region
\be
\label{rcexp}
R_{i}^4=2 \mu_i\cdot \mu_i
\hskip 1in
A_i^2 =  {16 \kappa_i\cdot \kappa_i\over \mu_i\cdot \mu_i}  
\ee
From  the leading terms (\ref{asymmeta})  we deduce that the metric takes an asymptotic 
$AdS_3\times S^3$ form. The subleading terms depend in general on the spherical slicing 
coordinate $\theta$. In the Fefferman-Graham the metric takes the following form,
\bea
\label{adsmet2}
ds^2_{FG}\sim  R_i^2 \Big\{ {1\over u^2} \Big( du^2 - g_1d t^2+ g_2    dx_{\perp}^2 \Big) 
+ g_3 \Big( d y^2+ \omega   dx_\perp\Big)^2 + g_4  \sin^2 y ds_{S^2}^2\Big\}
 \eea
Here $u$ is a Poincar\'e slicing coordinate where  $u= 0$ corresponds to the boundary of 
$AdS$ and the UV cutoff is defined by restricting the range  $u>\ep$. The coordinate $x_\perp$ 
denotes the distance from the junction and has a range $x_\perp\in [0,\infty ]$.  
Scaling symmetry implies that the functions $g_k$ and $ \omega$ depend only on $y$ and 
the combination $x_\perp/ u$. In the limit $u\to 0$ the $g_k$ behave as,
\be
\lim_{u\to 0} g_{k} \left ( {x_\perp\over u} , y \right ) =1, \;\; {k=1,2,3,4} 
\quad \quad \lim_{u\to 0} \omega \left ( {x_\perp\over u} , y \right ) =0
\ee
The maps between the $AdS_2$ slicing (\ref{adsmet1}) and the Fefferman-Graham 
coordinate slicing (\ref{adsmet2}) can be constructed in an asymptotic expansion in   $e^{-x}$.
\bea
\label{mapa}
{u\over z} &=& {2\over A_i} e^{-x} + k_1(\theta) e^{-2x}+\cdots
\nonumber\\
{x_\perp\over z} &=& 1+  l_1(\theta)   e^{-x} +\cdots\nonumber\\
y &=& \theta+  m_1(\theta) e^{-x} +\cdots
\eea
As discussed  \cite{Papadimitriou:2004rz,Jensen:2013lxa,Estes:2014hka} this coordinate map 
breaks down when $x_\perp$ becomes small compared to $u$. The region where the map is 
applicable is called a ``Fefferman-Graham" (FG) patch". For an N-junction all FG patches are 
smoothly joined to a central patch when  $x_\perp\sim u$.  While  it is very hard to find such a 
map (see  \cite{Jensen:2013lxa,Estes:2014hka,Papadimitriou:2004rz} for a discussion) we note 
that for the identification of the cutoff only map in the FG patch is needed since we can choose 
the location of the entangling surface   $x_\perp=L$ to be much larger than the UV cutoff $u=\ep$. 
It follows from (\ref{mapa}) that in this case $x$ is large and we are safely in the FG patch.

\sm

The relation of the UV cutoff $\ep$ and the radial integration cutoff in $\ep_i$ in the $i$-th 
asymptotic $AdS_3 \times S^3$ region is given by,
\bea
\ep_i &=& e^{-x^{(i)}_{\ep}}= A_i \, {\ep\over 2L} 
\eea
Plugging in the expression for $A_i$ give in (\ref{rcexp}) and squaring one arrives at,
\be
\ep_i^2 = {4 \kappa_i \cdot \kappa_i \over \mu_i \cdot \mu_i} \left({\ep\over L} \right)^2
\ee
Hence we have arrived  at (\ref{3a2}).



\newpage
 
 


 
\begin{thebibliography}{99}
\itemsep=0.0in

\bibitem{Maldacena:1997re}
  J.~M.~Maldacena,
  ``The Large N limit of superconformal field theories and supergravity,''
  Adv.\ Theor.\ Math.\ Phys.\  {\bf 2} (1998) 231
  [hep-th/9711200].


\bibitem{Gubser:1998bc}
  S.~S.~Gubser, I.~R.~Klebanov and A.~M.~Polyakov,
  ``Gauge theory correlators from noncritical string theory,''
  Phys.\ Lett.\ B {\bf 428} (1998) 105
  [hep-th/9802109].

\bibitem{Witten:1998qj}
  E.~Witten,
  ``Anti-de Sitter space and holography,''
  Adv.\ Theor.\ Math.\ Phys.\  {\bf 2} (1998) 253
  [hep-th/9802150].

\bibitem{Ryu:2006bv}
  S.~Ryu and T.~Takayanagi,
  ``Holographic derivation of entanglement entropy from AdS/CFT,''
  Phys.\ Rev.\ Lett.\  {\bf 96} (2006) 181602
  [hep-th/0603001].

\bibitem{Ryu:2006ef}
  S.~Ryu and T.~Takayanagi,
  ``Aspects of Holographic Entanglement Entropy,''
  JHEP {\bf 0608} (2006) 045
  [hep-th/0605073].


\bibitem{Affleck:1991tk}
  I.~Affleck and A.~W.~W.~Ludwig,
  ``Universal noninteger 'ground state degeneracy' in critical quantum systems,''
  Phys.\ Rev.\ Lett.\  {\bf 67} (1991) 161.

\bibitem{Karch:2000gx}
  A.~Karch and L.~Randall,
  ``Open and closed string interpretation of SUSY CFT's on branes with boundaries,''
  JHEP {\bf 0106} (2001) 063
  [hep-th/0105132].

\bibitem{Bachas:2001vj}
  C.~Bachas, J.~de Boer, R.~Dijkgraaf and H.~Ooguri,
  ``Permeable conformal walls and holography,''
  JHEP {\bf 0206} (2002) 027
  [hep-th/0111210].

\bibitem{Bachas:2007td}
  C.~Bachas and I.~Brunner,
  ``Fusion of conformal interfaces,''
  JHEP {\bf 0802} (2008) 085
  [arXiv:0712.0076 [hep-th]].

\bibitem{Fuchs:2007tx}
  J.~Fuchs, M.~R.~Gaberdiel, I.~Runkel and C.~Schweigert,
  ``Topological defects for the free boson CFT,''
  J.\ Phys.\ A {\bf 40} (2007) 11403
  [arXiv:0705.3129 [hep-th]].


\bibitem{Clark:2004sb}
  A.~B.~Clark, D.~Z.~Freedman, A.~Karch and M.~Schnabl,
  ``Dual of the Janus solution: An interface conformal field theory,''
  Phys.\ Rev.\ D {\bf 71} (2005) 066003
  [hep-th/0407073].

\bibitem{D'Hoker:2006uv}
  E.~D'Hoker, J.~Estes and M.~Gutperle,
  ``Interface Yang-Mills, supersymmetry, and Janus,''
  Nucl.\ Phys.\ B {\bf 753} (2006) 16
  [hep-th/0603013].
  
\bibitem{Gaiotto:2008sa}
  D.~Gaiotto and E.~Witten,
  ``Supersymmetric Boundary Conditions in N=4 Super Yang-Mills Theory,''
  J.\ Statist.\ Phys.\  {\bf 135} (2009) 789
  [arXiv:0804.2902 [hep-th]].

\bibitem{Gaiotto:2008sd}
  D.~Gaiotto and E.~Witten,
  ``Janus Configurations, Chern-Simons Couplings, And The theta-Angle in N=4 Super Yang-Mills Theory,''
  JHEP {\bf 1006} (2010) 097
  [arXiv:0804.2907 [hep-th]].
 
\bibitem{Gaiotto:2009fs}
  D.~Gaiotto,
  ``Surface Operators in N = 2 4d Gauge Theories,''
  JHEP {\bf 1211} (2012) 090
  [arXiv:0911.1316 [hep-th]].
 


\bibitem{Drukker:2010jp}
  N.~Drukker, D.~Gaiotto and J.~Gomis,
  ``The Virtue of Defects in 4D Gauge Theories and 2D CFTs,''
  JHEP {\bf 1106} (2011) 025
  [arXiv:1003.1112 [hep-th]].
 
\bibitem{Bak:2003jk}
  D.~Bak, M.~Gutperle and S.~Hirano,
  ``A Dilatonic deformation of AdS(5) and its field theory dual,''
  JHEP {\bf 0305} (2003) 072
  [hep-th/0304129].


\bibitem{Clark:2005te}
  A.~Clark and A.~Karch,
  ``Super Janus,''
  JHEP {\bf 0510} (2005) 094
  [hep-th/0506265].
  

\bibitem{D'Hoker:2007xy}
  E.~D'Hoker, J.~Estes and M.~Gutperle,
  ``Exact half-BPS Type IIB interface solutions. I. Local solution and supersymmetric Janus,''
  JHEP {\bf 0706} (2007) 021
  [arXiv:0705.0022 [hep-th]].

\bibitem{Lunin:2006xr}
  O.~Lunin,
  ``On gravitational description of Wilson lines,''
  JHEP {\bf 0606} (2006) 026
  [hep-th/0604133].

\bibitem{D'Hoker:2007fq}
  E.~D'Hoker, J.~Estes and M.~Gutperle,
  ``Gravity duals of half-BPS Wilson loops,''
  JHEP {\bf 0706} (2007) 063
  [arXiv:0705.1004 [hep-th]].


\bibitem{D'Hoker:2008wc}
  E.~D'Hoker, J.~Estes, M.~Gutperle and D.~Krym,
  ``Exact Half-BPS Flux Solutions in M-theory. I: Local Solutions,''
  JHEP {\bf 0808} (2008) 028
  [arXiv:0806.0605 [hep-th]].

\bibitem{D'Hoker:2007xz}
  E.~D'Hoker, J.~Estes and M.~Gutperle,
  ``Exact half-BPS Type IIB interface solutions. II. Flux solutions and multi-Janus,''
  JHEP {\bf 0706} (2007) 022
  [arXiv:0705.0024 [hep-th]].

\bibitem{Lunin:2007ab}
  O.~Lunin,
  ``1/2-BPS states in M theory and defects in the dual CFTs,''
  JHEP {\bf 0710} (2007) 014
  [arXiv:0704.3442 [hep-th]].


\bibitem{Chiodaroli:2009xh}
  M.~Chiodaroli, E.~D'Hoker and M.~Gutperle,
  ``Open Worldsheets for Holographic Interfaces,''
  JHEP {\bf 1003} (2010) 060
  [arXiv:0912.4679 [hep-th]].

\bibitem{Chiodaroli:2010mv}
  M.~Chiodaroli, M.~Gutperle, L.~-Y.~Hung and D.~Krym,
  ``String Junctions and Holographic Interfaces,''
  Phys.\ Rev.\ D {\bf 83} (2011) 026003
  [arXiv:1010.2758 [hep-th]].
  
\bibitem{Chiodaroli:2011nr}
  M.~Chiodaroli, E.~D'Hoker, Y.~Guo and M.~Gutperle,
  ``Exact half-BPS string-junction solutions in six-dimensional supergravity,''
  JHEP {\bf 1112} (2011) 086
  [arXiv:1107.1722 [hep-th]].
  
\bibitem{Chiodaroli:2011fn}
  M.~Chiodaroli, E.~D'Hoker and M.~Gutperle,
  ``Simple Holographic Duals to Boundary CFTs,''
  JHEP {\bf 1202} (2012) 005
  [arXiv:1111.6912 [hep-th]].






\bibitem{Bachas:2013nxa}
  C.~P.~Bachas, I.~Brunner, M.~R.~Douglas and L.~Rastelli,
  ``Calabi's diastasis as interface entropy,''
  arXiv:1311.2202 [hep-th].
  

\bibitem{calabi}E. Calabi,``Isometric Imbedding of Complex Manifolds",  Ann. Math. {\bf 58}, 1 (1953).



\bibitem{Romans:1986er}
  L.~J.~Romans,
  ``Selfduality for Interacting Fields: Covariant Field Equations for Six-dimensional Chiral Supergravities,''
  Nucl.\ Phys.\ B {\bf 276} (1986) 71.


\bibitem{Lashkari:2013koa}
  N.~Lashkari, M.~B.~McDermott and M.~Van Raamsdonk,
  ``Gravitational dynamics from entanglement 'thermodynamics',''
  JHEP {\bf 1404} (2014) 195
  [arXiv:1308.3716 [hep-th]].

\bibitem{Faulkner:2013ica}
  T.~Faulkner, M.~Guica, T.~Hartman, R.~C.~Myers and M.~Van Raamsdonk,
  ``Gravitation from Entanglement in Holographic CFTs,''
  JHEP {\bf 1403} (2014) 051
  [arXiv:1312.7856 [hep-th]].



\bibitem{Azeyanagi:2007qj}
  T.~Azeyanagi, A.~Karch, T.~Takayanagi and E.~G.~Thompson,
  ``Holographic calculation of boundary entropy,''
  JHEP {\bf 0803} (2008) 054
  [arXiv:0712.1850 [hep-th]].
 
\bibitem{Brown:1986nw}
  J.~D.~Brown and M.~Henneaux,
  ``Central Charges in the Canonical Realization of Asymptotic Symmetries: An Example from Three-Dimensional Gravity,''
  Commun.\ Math.\ Phys.\  {\bf 104} (1986) 207.

\bibitem{Chiodaroli:2010ur}
  M.~Chiodaroli, M.~Gutperle and L.~-Y.~Hung,
  ``Boundary entropy of supersymmetric Janus solutions,''
  JHEP {\bf 1009} (2010) 082
  [arXiv:1005.4433 [hep-th]].
  
  
  
\bibitem{Chiodaroli:2012vc}
  M.~Chiodaroli, E.~D'Hoker and M.~Gutperle,
  ``Holographic duals of Boundary CFTs,''
  JHEP {\bf 1207} (2012) 177
  [arXiv:1205.5303 [hep-th]].
   
     
\bibitem{Holzhey:1994we}
  C.~Holzhey, F.~Larsen and F.~Wilczek,
  ``Geometric and renormalized entropy in conformal field theory,''
  Nucl.\ Phys.\ B {\bf 424} (1994) 443
  [hep-th/9403108].
   
\bibitem{Calabrese:2004eu}
  P.~Calabrese and J.~L.~Cardy,
  ``Entanglement entropy and quantum field theory,''
  J.\ Stat.\ Mech.\  {\bf 0406} (2004) P06002
  [hep-th/0405152].
  
    
     
\bibitem{Jensen:2013lxa} 
  K.~Jensen and A.~O'Bannon,
  ``Holography, Entanglement Entropy, and Conformal Field Theories with Boundaries or Defects,''
  Phys.\ Rev.\ D {\bf 88}, 106006 (2013)
  [arXiv:1309.4523 [hep-th]].
  
\bibitem{Estes:2014hka}
  J.~Estes, K.~Jensen, A.~O'Bannon, E.~Tsatis and T.~Wrase,
  ``On Holographic Defect Entropy,''
  JHEP {\bf 1405} (2014) 084
  [arXiv:1403.6475 [hep-th]].
  
\bibitem{Papadimitriou:2004rz}
  I.~Papadimitriou and K.~Skenderis,
  ``Correlation functions in holographic RG flows,''
  JHEP {\bf 0410} (2004) 075
  [hep-th/0407071].


%
%
\end{thebibliography}
\end{document}